\input harvmac
\input epsf

\overfullrule=0pt
\abovedisplayskip=12pt plus 3pt minus 1pt
\belowdisplayskip=12pt plus 3pt minus 1pt
%
\def\tilde{\widetilde}
\def\bar{\overline}

\def\cN{{\cal N}}

\def\bigone{\hbox{1\kern -.23em {\rm l}}}
\def\ZZ{\hbox{\zfont Z\kern-.4emZ}}
\def\half{{\litfont {1 \over 2}}}

\def\tr{{\rm tr}\,}

\font\litfont=cmr6

\def\vx{{\vec x}}
\def\vy{{\vec y}}
\def\vX{{\vec X}}
\def\vY{{\vec Y}}
\def\vzero{{\vec 0}}
\def\dbar#1{{\bar {\rm D}#1}}
\def\mofl{(-1)^{F_L}}

\def\ddb#1{{${\rm D}#1-{\bar{\rm D}#1}$}}
\def\Dbar{${\bar {\rm D}}$}
\def\onebar{{\bar 1}}

\def\crr#1{~{C_{RR}^{(#1)}}}
\def\sign{{\rm sign}\,}
\def\Dp{D$p$}
\def\tq{{\tilde q}}
\def\tick{{\scriptstyle\surd\,}}

\lref\senreview{A. Sen, {\it ``Non-BPS States and Branes in String
Theory''}, hep-th/9904207.}
\lref\senworld{A. Sen, {\it ``Supersymmetric World-Volume Action For
Non-BPS D-Branes''}, hep-th/9909062; JHEP {\bf 10} (1999) 008.} 
\lref\sentach{A. Sen, {\it ``Tachyon Condensation on the Brane
Anti-Brane System''}, hep-th/9805170; JHEP {\bf 08} (1998) 012.}
\lref\senuniv{A. Sen, {\it ``Universality of the Tachyon Potential''}, 
hep-th/9911116.}
\lref\senbound{A. Sen, {\it ``Stable Non-BPS Bound States of 
BPS D-branes''}, hep-th/9805019; JHEP {\bf 08} (1998) 010.}
\lref\sencycle{A. Sen, {\it ``BPS D-Branes on Nonsupersymmetric
Cycles''}, hep-th/9812031; JHEP {\bf 12} (1998) 021.}
\lref\joysen{J. Majumder and A. Sen, {\it `` `Blowing Up' D-Branes 
on Nonsupersymmetric Cycles''}, hep-th/9906109; JHEP {\bf 09} (1999)
004.} 
\lref\gabsen{M. Gaberdiel and A. Sen, {\it ``Nonsupersymmetric
D-Brane Configurations with Bose-Fermi Degenerate Open String
Spectrum''}, hep-th/9908060; JHEP {\bf 11} (1999) 008.}
\lref\naraintach{E. Gava, K.S. Narain and M.H. Sarmadi, {\it ``On the 
Bound States of p-Branes and (p+2)-Branes''}, hep-th/9704006;
Nucl. Phys. {\bf B504} (1997) 214.}
\lref\witk{E. Witten, {\it ``D-branes and K-theory''},
hep-th/9810188; JHEP {\bf 12} (1998) 019.}
\lref\hork{P. Horava, {\it ``Type IIA D-Branes, K-Theory and Matrix
Theory''}, hep-th/9812135; Adv. Theor. Math. Phys. {\bf 2} (1999)
1373.}
\lref\piljin{P. Yi, {\it ``Membranes from Five-Branes and Fundamental
Strings from Dp-Branes''}, hep-th/9901159; 
Nucl. Phys. {\bf B550} (1999) 214.}
\lref\mons{N. Manton, {\it ``Monopole Interactions at Long Range''}, 
Phys. Lett. {\bf B154} (1985) 397; 
(E) {\bf B157} (1985) 475\semi
N. Manton and G. Gibbons,  {\it ``The Moduli Space for Well-Separated 
BPS Monopoles''}, hep-th/9506052;  Phys. Lett. {\bf B356} (1995) 32\semi
K. Lee, E. Weinberg and P. Yi, {\it ``The Moduli Space of Many BPS 
Monopoles for Arbitrary Gauge Groups''}, hep-th/9602167; Phys. Rev. 
{\bf D54} (1996) 1633.}
\lref\christim{C. Fraser and T. Hollowood, {\it ``Semi-classical 
Quantisation in N=4 Supersymmetric Yang-Mills Theory and Duality''}, 
hep-th/9704011; Phys. Lett. {\bf B402} (1997) 106.}
\lref\polkthree{J. Polchinski, {\it ``Tensors from K3 Orientifolds''},
hep-th/9606165; Phys. Rev. {\bf D55} (1997) 6423.}
\lref\dmfrac{K. Dasgupta and S. Mukhi, {\it ``Brane Constructions, 
Fractional Branes and Anti-de Sitter Domain Walls''}, hep-th/9904131;
JHEP {\bf 07} (1999) 008.}
\lref\dmconif{K. Dasgupta and S. Mukhi, {\it ``Brane Constructions,
Conifolds and M-Theory''}, hep-th/9811139; Nucl. Phys. {\bf B551} 
(1999) 204.}
\lref\urangaconif{A. Uranga, {\it ``Brane Configurations for Branes at
Conifolds''}, hep-th/9811004; JHEP {\bf 01} (1999) 022.}
\lref\witfourd{E. Witten, {\it ``Solutions of Four Dimensional Field
Theories Via M-theory''}, hep-th/9703166; Nucl. Phys. {\bf B500}
(1997) 3.}
\lref\hanwit{A. Hanany and E. Witten, {\it ``Type IIB Superstrings, 
BPS Monopoles, and Three-dimensional Gauge Dynamics''},
hep-th/9611230; Nucl. Phys. {\bf B492} (1997) 152.}
\lref\dougmoore{M. Douglas and G. Moore, {\it ``D Branes, Quivers and ALE
Instantons''}, hep-th/9603167.}
\lref\fracbranes{M. Douglas, {\it ``Enhanced Gauge Symmetry in M(atrix)
Theory''}, hep-th/9612126; JHEP {\bf 07} (1997) 004\semi 
D.-E. Diaconescu, M. Douglas and J. Gomis, {\it ``Fractional Branes and 
Wrapped Branes''}, hep-th/9712230; JHEP {\bf 02} (1998) 013.} 
\lref\nsdual{H. Ooguri and C. Vafa, {\it ``Two-Dimensional Black Hole
and Singularities of CY Manifolds''}, hep-th/9511164; Nucl. Phys. 
{\bf B463} (1996) 55\semi
B. Andreas, G. Curio and D. L\"ust, {\it ``The Neveu-Schwarz 
Five-Brane and its Dual Geometries''}, hep-th/9807008; 
JHEP {\bf 10} (1998) 022\semi
A. Karch, D. L\"ust and D. Smith, {\it ``Equivalence of 
Geometric Engineering and Hanany-Witten via Fractional Branes''},
hep-th/9803232; Nucl. Phys. {\bf B533} (1998) 348.}
\lref\bergabrev{O. Bergman and M. Gaberdiel, {\it ``Non-BPS Dirichlet
Branes''}, hep-th/9908126.}
\lref\kenwilk{C. Kennedy and A. Wilkins, {\it ``Ramond-Ramond
Couplings on Brane-Antibrane Systems''}, hep-th/9905195;
Phys. Lett. {\bf B464} (1999) 206.}
\lref\gabstef{M. Gaberdiel and B. Stefanski, {\it ``Dirichlet Branes on
Orbifolds''}, hep-th/9910109.}
\lref\kinky{E. Bergshoeff and P.K. Townsend, {\it ``Solitons on the 
Supermembrane''}, hep-th/9904020; JHEP {\bf 9905} (1999) 021\semi
N. Lambert and D. Tong, {\it ``Kinky D-Strings''}, hep-th/9907098; 
to appear in Nucl. Phys. B.}
\lref\strass{M. Strassler, {\it ``Confining Phase of Three Dimensional 
Supersymmetric Quantum Electrodynamics''}, hep-th/9912142.}
\lref\poly{A.M. Polyakov, {\it ``Quark Confinement and Topology of Gauge 
Groups''}, Nucl. Phys. {\bf B120} (1977) 429.} 
\lref\aldauranga{G. Aldazabal and A.M. Uranga, {\it ``Tachyon Free
Nonsupersymmetric Type IIB Orientifolds Via Brane-AntiBrane 
Systems''}, hep-th/9908072; JHEP {\bf 10} (1999) 024.}
\lref\tatar{M. Mihailescu, K. Oh and R. Tatar, {\it ``Non-BPS Branes 
on a Calabi-Yau Threefold and Bose-Fermi Degeneracy''}, hep-th/9910249.}
\lref\lerda{A. Lerda and R. Russo, {\it ``Stable Non-BPS States in 
String Theory: A Pedagogical Review''}, hep-th/9905006.} 
\lref\gaunthull{J.P. Gauntlett and C.M. Hull, {\it ``BPS 
States With Extra Supersymmetry''}, hep-th/9909098; JHEP {\bf 01}
(2000) 004.}

{\nopagenumbers
\Title{\vbox{
\hbox{hep-th/0001066}
\hbox{TIFR/TH/00-02}
\hbox{KCL-TH-00-02}}}
{\centerline{Brane-Antibrane Constructions}}
\vskip -12pt
\centerline{{Sunil Mukhi}$^1$\footnote{}{E-mail: mukhi@tifr.res.in,
nemani@tifr.res.in, tong@mth.kcl.ac.uk}, {Nemani
V. Suryanarayana}$^1$ and {David Tong}$^2$}
\vskip 8pt
\centerline{\it ${}^1$ Tata Institute of Fundamental Research,}
\centerline{\it Homi Bhabha Rd, Mumbai 400 005, India}
\vskip 5pt
\centerline{\it ${}^2$ Department of Mathematics, Kings College,}
\centerline{\it The Strand, London, WC2R 2LS, UK}

\vskip 15pt
\centerline{ABSTRACT}

In type II string theories, we examine intersecting brane
constructions containing brane-antibrane pairs suspended between
5-branes, and more general non-BPS constructions. The tree-level
spectra are obtained in each case. We identify various models with
distinct physics: parallel brane-antibrane pairs, adjacent pairs,
non-adjacent pairs, and configurations which break all supersymmetry
even though any pair of branes preserves some supersymmetry. In each
case we examine the possible decay modes. Some of these configurations
turn out to be tachyon-free, stable non-BPS states. We use T-duality
to map some of our brane constructions to brane-antibrane pairs at ALE
singularities. This enables us to explicitly derive the spectra by the
analogue of the quiver construction, and to compute the sign of the
brane-antibrane force in each case.
\vfill
\Date{January 2000}
\eject}
\ftno=0

\newsec{Introduction}

The study of unstable branes in type II superstring theories has made
considerable progress over the last two years\refs\senreview. The
relevant unstable branes are of two types: \Dp-brane-antibrane pairs,
($p$ even for type IIA, odd for type IIB) and unstable \Dp-branes with
$p$ odd for type IIA, even for type IIB. In both cases, the
instability is signalled by the presence of a tachyon, and it is
possible to identify a variety of decay modes. These assemble
themselves into interesting sequences that terminate with stable BPS
D-branes.

The unstable branes and their decay modes form a beautiful and
fundamental structure, which has been interpreted in terms of
K-theory\refs{\witk,\hork}. This structure can then be used to study
more complicated situations such as orientifolds, orbifolds and
Calabi-Yau compactifications. In these cases it often happens that one
discovers novel stable non-BPS states (see
Refs.\refs{\senreview,\bergabrev,\lerda} and references therein).

In the present paper, we attempt to generalise these elegant
discoveries to situations where unstable D-branes are suspended
between other branes. As we will see, along with many phenomena that
are familiar from the study of infinite or toroidally compactified
unstable branes, there are also new constraints and novel physical
situations that have no counterpart in the simpler models.

All the brane constructions that we study will have completely broken
supersymmetry. However, they are states of type II string theories,
and therefore are endowed with special properties that arise from the
fact that the underlying theory is supersymmetric. One interesting
class of models that we define has broken supersymmetry despite the
absence of brane-antibrane pairs or single unstable branes. In these
models one arranges at least three different types of branes together,
in such a way that each pair preserves some supersymmetry, but all the
branes together break all supersymmetry.

Even in supersymmetric models, the study of suspended branes suffers
from some uncertainties, as it is often hard to reliably extract the
spectrum of light states. These uncertainties are all the more severe
in the present case, as the brane configurations lack supersymmetry
which would have classified the possible states into multiplets. As
most of the models we study involve D-branes suspended between
NS5-branes, we will find it useful to dualise away the NS5-branes into
ALE geometry\refs\nsdual\ and then study the model using quiver
techniques\refs\dougmoore, where many of the relevant quantities can
in fact be reliably computed.

Although this paper will deal for the most part with unstable
configurations, it is meaningful to study their spectra at weak
coupling by working at string tree-level. Here one can identify the
presence or absence of potential decay modes related to tachyon
condensation, even though the configuration actually gets destabilised
after loop effects are taken into account. All discussions of unstable
configurations should be understood in the light of this comment.

The plan of this paper is as follows. In Section 2, we describe
systems of parallel D-brane-antibrane pairs suspended between
NS5-branes and D5-branes. These are closest in behaviour to the
noncompact parallel brane-antibrane pairs, though we identify some
differences that arise due to boundary conditions at the ends of the
D-branes. In Section 3 we look at unstable uncharged D-branes
suspended between NS5-branes and examine some possible decay modes. In
Section 4 we consider systems of parallel NS5-branes with a D-brane
stretched across one segment and a \Dbar-brane stretched across the
adjacent segment. In this case the pair cannot annihilate. We examine
some related models and argue that in general these pairs will repel
each other. We also consider configurations with more NS5-branes,
where the brane-antibrane pair is non-adjacent. In Section 5 we
introduce the ``Borromean branes'', configurations of branes in which
supersymmetry is broken by all the branes together but not by any
given pair of branes. This has the interesting consequence that there
is no perturbative tachyon, and the configuration can potentially be
stable. In Section 6 we describe the duality which relates suspended
branes between NS5-branes to fractional branes at ALE
singularities. Although this duality has been described and used
before in the literature, we give a slightly different and very
explicit derivation, which will hopefully make it somewhat clearer. In
Section 7 we apply this duality to analyse parallel and adjacent
brane-antibrane pairs from the point of view of quiver theory. In
Section 8 we compute the spectra of open strings in these
configurations by constructing boundary states for the relevant
fractional branes. We also use this formalism to compute the forces
between different pairs of fractional branes, confirming some of the
speculations made in earlier sections.

We will not provide a review of various fundamental aspects of
brane-antibrane dynamics that will be made use of in this paper. For
this, the reader is advised to consult Refs.\refs{\senreview,
\bergabrev,\lerda}.

\newsec{Suspended Parallel Brane-antibrane Pairs}

Consider, in type IIA string theory, a model with a pair of NS5-branes
extending along the directions $(x^1,x^2,x^3,x^4,x^5)$, and located at
$(x^6,x^7,x^8,x^9)=(0,0,0,0)$ and $(L_6,0,0,0)$ respectively. Thus
they are parallel and separated by a finite distance $L_6$ along
$x^6$. Between these, we suspend a D4-brane and a $\dbar4$-brane along
the $x^6$ direction. They extend along the directions $(x^1,x^2,x^3)$
and can be separated from each other along $(x^4,x^5)$
(Fig.(2.1)). Without the antibrane, this model would belong to the 
class of brane constructions analysed in Ref.\refs\witfourd.
\bigskip

\centerline{\epsfbox{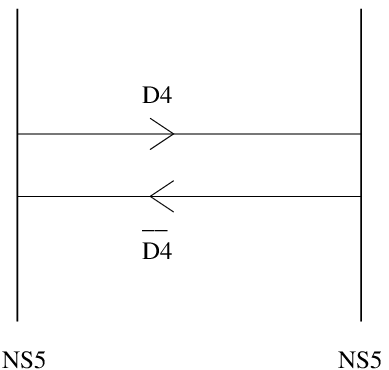}}\nobreak
\centerline{\footnotefont
Fig.(2.1): A \ddb4\ pair suspended between parallel 
NS5-branes in type IIA.}
\bigskip
We expect that the common 3+1-dimensional world-volume supports a
non-supersymmetric field theory. The spectrum of light states,
including possible tachyons, can be deduced as follows. First consider
an infinitely extended D4-$\dbar4$ pair. As is
well-known\refs\senreview, the open-string states have four sectors,
corresponding to the Chan-Paton factors:
\eqn\cpfactors{
a=\pmatrix{1&0\cr0&0\cr},\quad
b=\pmatrix{0&0\cr0&1\cr},\quad
c=\pmatrix{0&1\cr0&0\cr},\quad
d=\pmatrix{0&0\cr1&0\cr} }
The first two sectors come from strings beginning and ending on the
same brane(antibrane). Together, these give a GSO-projected spectrum
of $U(1)\times U(1)$ vector multiplets, containing two 4+1-dimensional
gauge fields $A_\mu^{(I)}$, two sets of five massless scalar fields
$X_i^{(I)}$ and two sets of 4 massless Majorana fermions
$\chi_{A}^{(I)}$, where $i=1,\ldots,5$, $A=1,\ldots, 4$ and $I=1,2$.
The index $I$ here labels the Chan-Paton factors $a,b$.  In the other
two sectors, associated to open strings going from the brane to the
antibrane and vice-versa, there is a complex tachyon $T$ carrying
charges $(1,-1)$ under $U(1)\times U(1)$, and another set of massless
Majorana fermions $\lambda_A^{(I)}$. (This time the index $I$ labels
the Chan-Paton factors $c,d$.) These correspond to ``anti-GSO'' states
that arise because of the fact that the two branes carry opposite RR
charges.

Together, the infinite brane-antibrane pair breaks all 32
supersymmetries of the bulk theory. Hence we expect to find 32
massless Goldstone fermions (``goldstinoes'') on the world-volume of
this pair. Actually we have 64 massless fermions, from the
$\chi_A^{(I)}$ and $\lambda_A^{(I)}$. One way to identify which of
these are goldstinoes is to use the fact that when we quotient type
IIA string theory, with a \ddb4\ pair, by the symmetry
$\mofl$, we end up with type IIB string theory with an unstable
D4-brane in it. This unstable brane carries a real tachyon and
precisely 32 massless Majorana fermions. In this case it is clear that
all the fermions must be goldstinoes.

It follows that in the \ddb4\ pair, the goldstinoes must be those
fermions which survive the $\mofl$ quotient ($\mofl$ has no twisted
sectors on open-string states, so those fermions which survive the
quotient and are goldstinoes in the final theory must have been
goldstinoes before taking the quotient too). This quotient acts on the
Chan-Paton factors as conjugation by $\sigma_1$, hence the surviving
Chan-Paton factors on the unstable D4-brane are $1,\sigma_1$. It
follows that $\chi^+_A = \chi_A^{(1)} + \chi_A^{(2)}$ and
$\lambda^+_A = \lambda_A^{(1)} + \lambda_A^{(2)}$ are the goldstinoes
on the \ddb4\ pair.

Now let us go back to the model of a \ddb4\ pair suspended between
NS5-branes and work out the spectrum. Again we have a $U(1)\times
U(1)$ gauge theory, whose spectrum is a truncation of that on the
infinite \ddb4\ pair. The truncation arises from the boundary
conditions at the two ends of the 4-branes.

For the massless fields in the GSO-projected sector (associated to
Chan-Paton factors $1,\sigma_3$) the truncation is well-known: the
massless scalars that are collective coordinates for broken
translation invariance are projected out whenever they correspond to
directions along which the 4-branes cannot move. Since the NS5-branes
on which they end are located at fixed values of $x^7,x^8,x^9$, it
follows that the scalars $X^{(I)}_7,X^{(I)}_8,X^{(I)}_9$ are
projected out. Along with these, the gauge field component $A^{(I)}_6$
and half the associated fermions, say $\chi^{(I)}_3,\chi^{(I)}_4$ are
projected out. The result is a set of massless fields:
$A^{(I)}_\mu,X^{(I)}_i, \chi^{(I)}_A$ where $\mu=0,\ldots,3$,
$i=4,5$ and $A=1,2$. This is a pair of vector multiplets of $\cN=2$
supersymmetry in 3+1 dimensions.

In the anti-GSO sector, supersymmetry is clearly not available to
guide us in finding the spectrum that survives.  We shall work in the
limit in which the separation between NS5-branes is very much smaller
than the string length scale, ensuring that the higher tachyon
Kaluza-Klein modes have positive mass-squared and so, as for other
fields, we restrict attention to the constant modes.  It is plausible,
and we will confirm this later, that the complex tachyon survives the
boundary conditions. Recall that for an infinite \ddb4\ pair, the
tachyon is associated to the NS sector ground state for open strings
stretching between the brane and the antibrane. Thus we are claiming
that this ground state is not projected out by the boundary
conditions. It is physically evident that the \ddb4\ pair suspended
between parallel NS5-branes should be unstable, just as an infinite,
or wrapped, \ddb4\ pair would be. Hence we expect it to contain a
tachyon. Another fact that supports this conclusion is that the
$U(1)\times U(1)$ gauge fields under which the tachyon is charged do
survive the boundary conditions, as we have seen. In a later section
we will show explicitly in a T-dual version of this model that the
tachyon is indeed present.

Finally we turn to the anti-GSO sector fermions. In the case of
infinite \ddb4\ pairs, we had altogether 32 anti-GSO fermions
$\lambda^{(I)}_A$, of which 16 were goldstinoes. Now on the suspended
\ddb4\ pair we expect to find altogether 16 goldstinoes (because they
break all 16 of the supersymmetries that are preserved by the
NS5-branes). But so far we have found 8 goldstinoes $\chi^+_A$,
$A=1,2$ and another 8 non-goldstino fermions $\chi^-_A = \chi_A^{(1)}
- \chi_A^{(2)}$, $A=1,2$. The remaining 8 goldstinoes must therefore
be $\lambda^+_A$, $A=1,2$. The symmetry between $\lambda^{(1)}_A$ and
$\lambda^{(2)}_A$ then suggests that the combinations $\lambda^-_A =
\lambda_A^{(1)} -
\lambda_A^{(2)}$, $A=1,2$ also survive as (non-goldstino) massless
fermions. 

To summarize, we have a tachyonic $U(1)\times U(1)$ gauge theory on
the world-volume of the \ddb4\ pair, with a pair of gauge fields 
$A_\mu^{(I)}$, massless scalars $X_i^{(I)}$ and massless fermions
$\chi^\pm_A, \lambda^\pm_A$ of which the $+$ superscripts correspond
to goldstinoes. Finally there is a complex tachyon $T=T_1+iT_2$. 

Now we are in a position to ask how this configuration can decay. The
first and most elementary process is that the tachyon can go to its
minimum. In this case, the brane-antibrane pair annihilates
completely. While one of the $U(1)$ gauge symmetries gets Higgsed in
the process, the other $U(1)$ gets ``confined'' by the mechanism
discussed in Ref.\refs\senworld. At the same time, the value of the
potential at the tachyonic minimum, $V(T_0)$, is expected to cancel
the tension of the annihilating branes. Thus we end up with a BPS
configuration consisting of just a pair of parallel NS5-branes. Note
that both the brane tensions and the tachyon potential scale by a
common factor of $L_6$, the separation between the NS5-branes, hence
apart from this overall scale factor, we expect $V(T_0)$ to be
independent of the coupling constant of the 3+1 dimensional field
theory. This is consistent with a recent analysis showing that upto
such a factor the tachyon potential is universal, independent of the
background\refs\senuniv.

A less elementary decay mode would be condensation of a tachyonic
vortex\refs{\naraintach,\sentach}. For infinite or wrapped \ddb p\
pairs, the complex tachyon can condense in a topologically stable
vortex, while the relative $U(1)$ gauge field under which it is
charged carries a unit of magnetic flux. The result is a stable BPS
D$(p-2)$ brane. Looking at the field content of the world-volume
theory on the suspended \ddb4\ pair, we see that the same
configuration is allowed here, providing the resulting D2-brane is
extended in the $x^6$ direction. The configuration of Fig.(2.1) can
therefore decay into a configuration of a BPS D2-brane suspended
between parallel NS5-branes.

The picture of this process on the world-volume of the NS5-branes is
interesting. A D4-brane ends as a 3-brane in the NS5-brane
world-volume. This 3-brane behaves as a vortex since it has
co-dimension 2. The $\dbar4$ brane similarly behaves as an
antivortex. The process where the tachyon goes to its minimum
corresponds to simple annihilation of the vortex-antivortex pair. On
the other hand, the more complex process in which the tachyon
condenses as a vortex, corresponds to the 3-brane anti-3-brane
vortices annihilating into a 1-brane\foot{This discussion may be a
little confusing since there are two types of vortices involved. The
tachyonic vortex is localized in, say, the $(x^1,x^2)$ directions,
while the 3-brane vortices in the NS5-branes are localized in the
$(x^4,x^5)$ directions.}. This 1-brane in the NS5 world-volume is just
the end of a D2-brane.

Next let us examine a decay mode which would be allowed for infinite
or wrapped \ddb4\ pairs but turns out to be forbidden in the present
situation. It has been noted\refs\piljin\ that any \ddb p\ pair can
also decay by creating an {\it electric} flux on its
world-volume. In this case, the decay product is a {\it fundamental}
string (F-string). If this were possible, the configuration of
Fig.(2.1) would decay into a configuration where an F-string is
suspended between two NS5-branes. Clearly, this is impossible since
F-strings cannot end on NS5-branes. Hence we must show that such 
an electric flux  is prohibited in the world-volume theory discussed
above for Fig.(2.1).

This follows from the fact that the boundary conditions imposed by
NS5-branes project out the component $A^{(I)}_6$ from each of the two
5-dimensional gauge fields. On \ddb4\ pairs without boundaries, this
component would dualize to form part of a magnetic 2-form gauge
potential, under which tachyonic strings arising from stretched
D2-branes are charged (although this phenomenon is, of course,
inherently non-perturbative). These magnetic tachyons could then
condense, forming an {\it electric} world-volume flux\refs\piljin\
(the dual of the usual vortex condensation where an electric tachyon
condenses giving rise to a magnetic flux). This process corresponds to
the annihilation of the pair into a fundamental string. Once the
NS5-branes are put in as boundaries, components of the magnetic 3-form
field strength $H_{\mu\nu\rho}$ with $\mu,\nu,\rho\neq 6$ are lost,
along with the possibility of an electric flux in the $x^6$
direction. However, one is still left with the decay mode in which the
\ddb4\ pairs decay into a fundamental string oriented parallel to the
NS5-branes in either the $x^1,x^2$ or $x^3$ directions. As with
similar decays in the case of infinitely extended branes, the
resulting configuration preserves some fraction of supersymmetry
which, in the present situation, is $1/4$ of the original 32
supercharges.

The fact that the \ddb4\ pair cannot decay into a fundamental string 
stretched between NS5-branes can also be understood in terms of the theory 
on the NS5-brane
worldvolume. In this language we start with a 3-brane
vortex-antivortex pair.  This cannot annihilate into a point-like
object (representing the end of a fundamental string), since a
fundamental string carries $B_{NS,NS}$ charge and therefore its
endpoint must be an electrically charged particle -- but the NS5-brane
of type IIA string theory does not carry 1-form gauge fields, and
therefore it does not support electrical charges.

For our next example, consider a \ddb3\ pair suspended between
NS5-branes in type IIB string theory (Fig.(2.2)). The system is very
similar to the one above. The NS5-branes have the same locations as
before, while the 3-branes extend along $(x^1,x^2,x^6)$ and are
located at definite positions in $(x^3,x^4,x^5)$ and at the origin in
$(x^7,x^8,x^9)$. Again, without the antibrane this model would be a
familiar one --- it belongs to the class of brane constructions
analysed in Ref.\refs\hanwit.
\bigskip

\centerline{\epsfbox{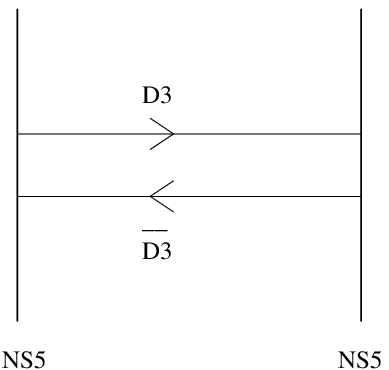}}\nobreak
\centerline{\footnotefont 
Fig.(2.2): A \ddb3\ pair suspended between parallel 
NS5-branes in type IIB.}
\bigskip
The world-volume theory now lives in 2+1 dimensions. The relevant
fields from the GSO-projected sector are a pair of $U(1)$ gauge
fields, a set of scalar triplets $X^{(I)}_i$, $i=3,4,5$, and a set of
massless fermions $\chi^{(I)}_A$, $A=1,\ldots,4$.  Note that this time
the fermions are two-component spinors. {}From the anti-GSO sector we
again expect a complex tachyon $T$ and a set of massless fermions
$\lambda^{(I)}_A$, $A=1,\ldots,4$. In this model, too, the
possibilities are that the \ddb3\ pair annihilates completely (with
the tachyon going to the minimum of the potential), or there is a
tachyonic vortex resulting in annihilation of the \ddb3\ pair into a
D-string stretched in the $x^6$ direction.  As in the model of
Fig.(2.1), the loss of $A_6$ due to boundary conditions ensures the
\ddb3\ pair can only decay into an F-string that lies parallel to the
NS5-branes.

In this model, it is natural to wonder whether more general decays to
(p,q)-strings occur. S-duality ensures that such a decay is allowed in
the case of infinitely extended \ddb3\ pairs, with the electric and
magnetic tachyons winding p and q times respectively before
condensing. Notice that the restrictions on the stable decay products
discussed above are precisely those of supersymmetry. The
corresponding supersymmetry restriction for a (p,q)-string is a
configuration in which the string ``kinks'' between the two
NS5-branes\refs\kinky. Unfortunately, we do not have enough
understanding of the non-perturbative magnetic tachyon dynamics to say
whether such a decay actually occurs.

The third model that we want to consider is quite different. Here we
have a \ddb3\ pair suspended between two D5-branes (Fig.(2.3)).
\bigskip

\centerline{\epsfbox{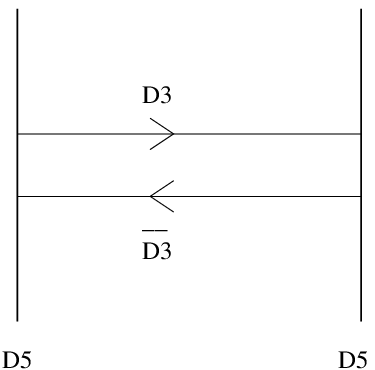}}\nobreak
\centerline{\footnotefont 
Fig.(2.3): A \ddb3\ pair suspended between parallel 
D5-branes in type IIB.}
\bigskip
In this case, the gauge field component $A_6$ is retained while the
components $A_\mu$, $\mu=0,1,2$ are projected out\refs\hanwit. The
scalar $A_6$ may be shifted by an integer multiple of $2\pi /L_6$ by
performing a ``large'' gauge transformation, which acts upon any
charged fields by shifting each Kaluza-Klein mode into the next. Thus
the scalar $A_6$ becomes periodic\foot{A well-known example of this
fact is that the moduli space of a D-string stretched between D3's is
the moduli space of a single monopole, which is $R^3 \times S^1$, with
periodicity $2\pi/L$.} and, as a result, this model has a {\it
magnetic} rather than electric gauge group. Although the model of
Fig.(2.3) is the S-dual of the model of Fig.(2.2), we cannot make any
definite use of this fact since we are dealing with unstable non-BPS
brane configurations. S-duality relates Fig.(2.2) at weak coupling to
Fig.(2.3) at strong coupling and vice-versa, while we are interested
in both models at weak coupling. The dynamics of the model of
Fig.(2.3) at strong coupling is rather clear: there is a magnetic
tachyon, corresponding to quantization of a D-string connecting the
\ddb3\ pair, and this can condense with an electric vortex to give an
F-string suspended between the D5-branes. This process is S-dual to
the condensation of an electric tachyon via a magnetic vortex in the
model of Fig.(2.2), and the resulting configuration (which is BPS) is
likewise S-dual to a D-string suspended between parallel NS5-branes.

The first statement that we can make about the model of Fig.(2.3) at 
weak coupling is 
that the boundary conditions imposed by the D5-branes on the \ddb3\ pair 
must forbid the decay of this pair into a D-string stretched between the 
two D5-branes. It is interesting to consider all the available
possibilities that lead to this conclusion. 

One possibility might be that the complex tachyon is projected out
along with the electric gauge field. In this case the \ddb3\ pair
suspended between two D5-branes would be stable for sufficiently close
D5-branes (since it is the constant mode of the tachyon that suffers
projection, one could still have a tachyonic instability for
well-separated D5-branes). This would mean that suspending a
\ddb3\ between D5-branes gives rise to a stable, non-BPS state for
some values of the separations. However, the present system is T-dual
to a D-string-anti-D-string pair suspended between parallel
D3-branes. In that system, the end-point of the D-string behaves as a
magnetic monopole in the D3-brane world-volume, so the
string-antistring pair is really like a monopole-antimonopole 
pair. Such a system can certainly annihilate to the vacuum and we 
therefore expect a tachyon to be present in the spectrum.

A second possibility is that, although the electric gauge field under 
which the tachyon was charged is projected out by the boundary conditions 
at the D5-brane, the complex tachyon itself survives. In this
scenario, the \ddb3\ pair can annihilate into the vacuum by having the
tachyon go to a constant minimum as usual. However, there also exists 
the possibility of the tachyon winding before condensing, potentially forming 
the forbidden configuration of a D-string stretched between D5-branes. 
One might hope that the resulting vortex solution has divergent 
energy since finite-energy vortices cannot exist in standard field
theories without a gauge field. In this case, we would not be able 
to identify the resulting configuration as a D-string. Moreover, the 
D-string charge coming from the Chern-Simons coupling 
$\int F\wedge C^{(2)}_{RR}$ would also be absent without a gauge field.

However, there are some reasons to be dissatisfied with this
scenario. Finite-energy vortices without gauge fields are possible if
one allows higher-derivative couplings in the field theory, which are
certainly present because of stringy corrections. Moreover, besides
the Chern-Simons term described above, there is another proposed
coupling\refs\kenwilk\ of the form $\int d(TDT^*)\wedge C^{(2)}_{RR}$
on a brane-antibrane pair, which could give rise to the induced
D-string charge without the help of a gauge field. Finally, it is
believed\refs\senreview\ that a \ddb3\ pair can decay by condensing a 
real {\it unstable} tachyonic kink, into an unstable D2-brane. In
turn, this can decay, by condensing a real stable tachyonic kink, into 
a BPS D-string. This two-step process could again lead to the forbidden
D-string, so (barring some mechanism that we do not presently
understand that inhibits one of the steps), the hypothesis that the
complex tachyon survives seems unlikely.

The final possibility is that one real component of the tachyon is
projected out, along with the electric gauge field. The surviving field
content of the suspended \ddb3\ pair would then be very similar to
that of a single unstable D-brane: a U(1) gauge field and scalars, and 
a neutral real tachyon. This has the advantage that the pair still has
two decay modes, out of the original three: it can decay into the
vacuum, by condensation of this real tachyon into its constant
minimum, or it can decay into the unstable D2-brane suspended between
D5-branes by condensation of a real tachyonic kink. Only vortex
condensation (whether in one step or two steps) is ruled out since 
there is only one tachyon.

This possibility seems the most elegant, because it inhibits only the
decay mode that is definitely forbidden on grounds of charge
conservation, into a D-string. Moreover, it suggests that an unstable
D2-brane of type IIB suspended between D5-branes is stable. We will
return to this point in the next section.

There is one remaining decay mode to consider in which the 
\ddb 3 pair decays  
into a D-string lying parallel to the D5-branes. Such a decay 
occurs if either a real or complex constant tachyonic mode 
survives and the topology required to realise these strings as solitonic 
solutions lies in the periodicity of $A_6$. The 3-dimensional field 
theory on the D3-branes includes the coupling $|A_6|^2|T|^2$ which 
ensures that when the tachyon condenses, the vacuum lies at $A_6=0$. 
However, $A_6$ may wind as, say, $x_1$ ranges from $-\infty$ to 
$+\infty$, resulting in a string like configuration stretched in the 
$x_2$ direction. This configuration has non-zero flux $F_{16}$ and so, 
by the usual Chern-Simons term, $\int F\wedge C^{(2)}_{RR}$, 
is identified as a D-string. 

There is another way to see the appearance of strings in this 
case\foot{We thank Kimyeong Lee and Piljin Yi for explanations 
of this point.} that dates back to the work of Polyakov \refs\poly. 
The periodic scalar, $A_6$, may be dualised in favour of  
a 3-dimensional gauge field, $\tilde{A}$. The tachyon couples 
magnetically to $\tilde{A}$ and, by the dual-Meissner effect, 
condensation of the tachyon leads to a vacuum in which objects charged 
electrically under $\tilde{A}$ are linearly confined. The electric 
flux lines associated with this confinement are then identified with 
the kinks described above. Similar systems, in which confining 
flux lines have a description in terms of classical solitons, have 
been considered recently in the supersymmetric context in \refs\strass.

For our final model, consider a \ddb3\ pair suspended between an 
NS5-brane at one end and a D5-brane at the other. The NS5-brane fills 
the directions
$(x^1,x^2,x^3,x^4,x^5)$ and is located at
$(x^6,x^7,x^8,x^9)=(0,0,0,0)$, while the D5-brane fills
$(x^1,x^2,x^7,x^8,x^9)$ and is located at
$(x^3,x^4,x^5,x^6)=(0,0,0,L_6)$. The \ddb3\ pair fills the directions
$(x^1,x^2)$, stretches from $x^6=0$ to $x^6=L_6$ and is located at
$(x^3,x^4,x^5,x^7,x^8,x^9)= (0,0,0,0,0,0)$ (Fig.(2.4)).
\bigskip

\centerline{\epsfbox{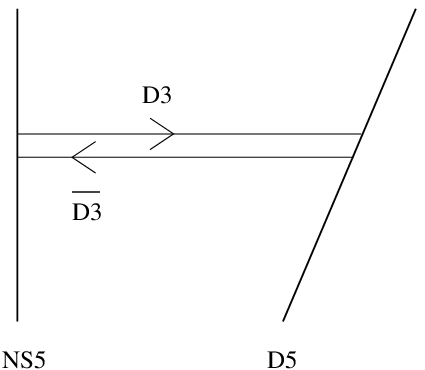}}\nobreak
\centerline{\footnotefont 
Fig.(2.4): A \ddb3\ pair suspended between an NS5
and a D5-brane in type IIB.}
\bigskip
{}From Ref.\refs\hanwit, we know that the massless states in the
GSO-projected sector, corresponding to open strings on the brane or
antibrane alone, get projected out at one or the other end so there is
neither an electric nor a magnetic gauge group. For the anti-GSO
sector, we again expect, as above, that either a real tachyon or the
whole complex tachyon survives at the D5 end. Whichever of these two
possibilities is realised will govern the dynamics of the system,
since the whole complex tachyon survives at the NS5-brane end. Note
that this model is self-dual under S-duality, so its strong-coupling
behaviour will be the same as its weak-coupling behaviour with the
roles of the two bounding 5-branes exchanged.

\newsec{Suspended Unstable D-branes}

The type IIA string has unstable, non-BPS \Dp-branes for
$p=1,3,5,7,9$\refs\senreview. Hence one can consider a brane
construction in which one of these is suspended between
NS5-branes. For example, let us start with an unstable D3-brane
suspended in this way (Fig.(3.1)).
\bigskip

\centerline{\epsfbox{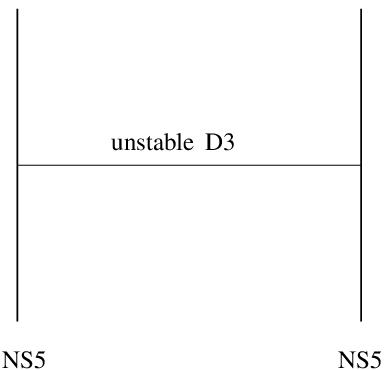}}\nobreak
\centerline{\footnotefont 
Fig.(3.1): An unstable D3-brane suspended between parallel 
NS5-branes in type IIA.}
\bigskip
The directions of the branes are exactly as for the model in
Fig.(2.2). Indeed, the configuration of Fig.(3.1) can be obtained by
starting with Fig.(2.2) and taking the quotient of the configuration
and the bulk theory together by $\mofl$. This is possible since the
brane construction in Fig.(2.2) is invariant under $\mofl$, which in
turn holds since $\mofl$ interchanges a D3-brane with a $\dbar3$-brane
in type IIB, and preserves the NS5-brane.

The spectrum on the world-volume theory of Fig.(3.1) is a truncation of
that on a single unstable D3-brane of type IIA. This unstable brane,
if it is infinite or wrapped, has a single gauge field $A_\mu$,
a set of massless scalars $X_i$ and a set of massless fermions
$\chi_A$, where $\mu=0,\ldots, 3$, $i=4,\ldots, 9$ and $A=1,\ldots,
4$. This is a GSO-projected spectrum and arises in the sector with
identity Chan-Paton factor. In addition, there is another sector, with
Chan-Paton factor $\sigma_1$ and anti-GSO projection, where one finds
a real tachyon and another set of massless fermions $\lambda_A$,
$A=1,\ldots, 4$. This time, all the 32 fermions $\chi_A$ and
$\lambda_A$ are goldstinoes of spontaneously broken
supersymmetry. Once the unstable brane is bounded by NS5-branes, it
breaks only 16 supersymmetries (the other 16 are already broken by the
NS5-branes which act as a ``background'' from the point of view of the
D3-brane). Thus we expect 16 massless fermions $\chi_A$ and
$\lambda_A$ where now $A=1,2$ and all these fermions are goldstinoes.
The rest of the light spectrum is made up by the gauge field $A_\mu$
and three scalars, $X_i$, $i=3,4,5$, along with a real tachyon
$T$.

There is another way to obtain the configuration of Fig.(3.1): start
with the configuration of Fig.(1.1), in type IIA theory, and allow the
real part of the (complex) tachyon on the \ddb4\ pair to develop a
kink. This is not a topologically stable solution, and therefore does
not cause the \ddb4\ to condense to a stable object. Instead it
creates an unstable D3-brane of type IIA at the point where the kink
is located\refs\senreview. To get the directions appropriate to
Fig.(3.1), the kink in Fig.(1.1) must be along the $x^3$ direction.

Since one way of obtaining Fig.(3.1) was to quotient the configuration
of Fig.(2.2) by $\mofl$, one is tempted to ask whether a similar
quotient can be carried out on Figs.(2.1) and (2.3). For Fig.(2.1)
this is indeed possible, and one ends up with the unstable D4-brane of
type IIB suspended between NS5-branes (Fig.(3.2)). However, for
Fig.(2.3) the story is quite different. The action of $\mofl$ does not
preserve D5-branes, hence it is not a symmetry of Fig.(2.3) and one cannot
quotient by it.
\bigskip

\centerline{\epsfbox{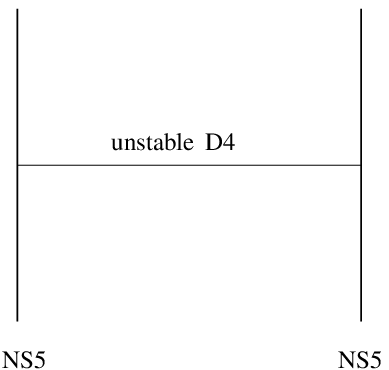}}\nobreak
\centerline{\footnotefont 
Fig. (3.2): An unstable D4-brane suspended between parallel 
NS5-branes in type IIB.}
\bigskip
Finally one can ask the reverse question: what happens when we
quotient Fig.(3.1) by $\mofl$? It is known that this action takes the
unstable D3-brane of type IIA to the stable BPS D3-brane of type
IIB. Hence it is reasonable to expect that quotienting Fig.(3.1) by
$\mofl$ gives us a single BPS D3-brane stretched between NS5-branes in 
type IIB theory. Likewise, we could take this quotient on Fig.(1.1) and we 
would end up with a stable D4-brane of type IIA suspended between two
NS5-branes. Thus, the stable BPS configurations analysed in
Refs.\refs{\witfourd,\hanwit}\ arise as $\mofl$ quotients of the
unstable configurations considered in this section.

Since unstable branes do not carry charges, one may ask if they can
always be consistently suspended between any pair of stable
branes. Consider a \ddb3\ pair suspended between NS5-branes in type
IIB. Clearly, condensation of a real tachyonic kink can give rise to
an unstable D2-brane suspended between the NS5-branes. The tachyon
that condenses is the real part of an electric tachyon on the \ddb3\
pair, which as we have argued, survives the boundary conditions
imposed by the NS5-brane. Thus it must be consistent to have an
unstable D2-brane suspended between NS5-branes, analogous to the
system in Fig.(3.2). (This configuration can also be produced by
starting with a \ddb2\ pair in type IIA, suspended between NS5-branes,
and quotienting by $\mofl$.) Moreover, the tachyon on this D2-brane,
although uncharged, is actually the imaginary part of the electric
tachyon on the original \ddb3\ pair. Thus, this tachyon is also not
projected out by the NS5-brane, and it can undergo further kink
condensation leading to a stable BPS configuration, a D-string
suspended between the NS5-branes.

If we replace the NS5-branes by D5-branes, we certainly cannot produce
such a configuration by starting from type IIA and quotienting by
$\mofl$, since type IIA has no BPS D5-branes. Thus the only way to
produce this configuration will be by kink condensation starting from
a \ddb3\ pair stretched between the D5-branes. This leads to two
possibilities, which are linked to the two possibilities considered in
the previous section. If the \ddb3\ pair retains a complex tachyon,
then condensation of a real kink will give rise to the unstable
D2-brane suspended between D5-branes. However, as noted in the
previous section, this does give rise to a potential paradox. The 
D2-brane would in turn be tachyonic, and its decay could potentially
produce a BPS D-string suspended between D5-branes, which we know to
be inconsistent. 

A more plausible scenario, in the light of our discussion in the
previous section, arises if the \ddb3\ pair between D5-branes retains
only one real tachyon. In that case, condensation of a tachyonic kink
would lead to a D2-brane suspended between D5-branes, which in turn
would be stable because there is no second tachyon. An amusing
feature of this scenario is that instead of the usual 3-step chain:
${\rm D}3-\dbar3\rightarrow {\rm D}2\rightarrow {\rm D}1$, where the
successive elements have two tachyons, one tachyon and no tachyon, we
find a shorter chain: $suspended~{\rm D}3-\dbar3
\rightarrow suspended~{\rm D}2$, where the two elements have
respectively one tachyon and no tachyon.

In this situation, note that the suspended D2-brane not only has no
2-brane charge (which is the defining property of an unstable D-brane)
but also no induced lower-form charges. Such charges can only arise
via world-volume Chern-Simons couplings involving a tachyon and
various even-form RR potentials, so once the tachyon is projected out,
this leaves a totally uncharged object.

\newsec{Suspended Adjacent Brane-Antibrane Pairs}

In this section we introduce three parallel NS5-branes in type IIA
theory and consider a system in which a D4-brane and a $\dbar4$-brane
end on a common NS5-brane from {\it opposite} sides. The configuration
is that of Fig.(4.1).
\bigskip

\centerline{\epsfbox{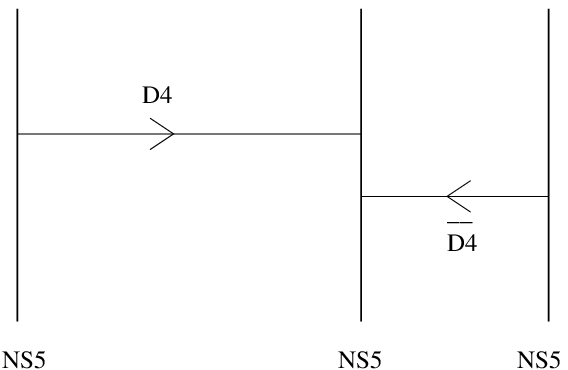}}\nobreak
\centerline{\footnotefont 
Fig.(4.1): Adjacent \ddb4\ pairs in type 
IIA.}
\bigskip
This type of configuration is much harder to analyse. Even for the BPS
analogue (with a D-brane on either side of an NS5-brane), it is
difficult to reliably extract the spectrum, and one has to use the
fact that one can continuously interpolate to a different
configuration of intersecting branes where perturbation theory is
reliable\refs\hanwit. Alternatively one can use duality, as we will do 
in subsequent sections.

The question we want to address is whether this configuration has an
instability, and if so, what is its nature and to what final
configuration does the system tend. On physical grounds, we might
expect the system to be tachyonic and unstable. If there is an
attractive force between the brane-antibrane pair, then they will tend
to line up as in Fig.(4.2).
\bigskip

\centerline{\epsfbox{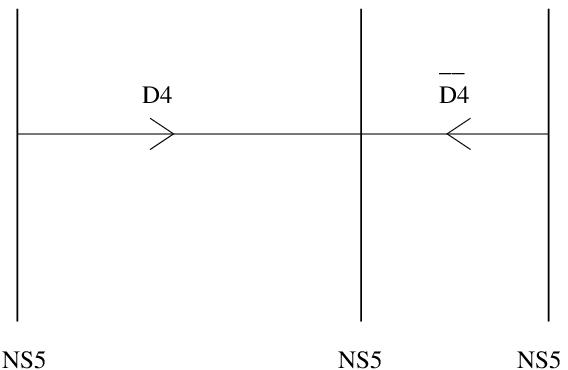}}\nobreak
\centerline{\footnotefont 
Fig.(4.2): Lining up of adjacent \ddb4\ pairs in type IIA.}
\bigskip
On the other hand, if the force is repulsive then they will move
apart. {}From the point of view of one of the pair, the other will go to 
infinity. 

Rather than immediately studying the above problem, we turn first to a 
related system that is more amenable to calculation: that of an 
adjacent \ddb1\ pair suspended between D3-branes. This configuration 
is related to the above by S- and T-dualities. However, the need to 
perform an S-duality, relating strong and weak coupling regimes, means 
that the physics of this system is not necessarily the same as that 
of Fig.(4.1). Nevertheless, in a later section we will return to 
the configuration of Fig.(4.1) and find the same behaviour. 

The D3-branes support an $\cN=4$, $d=3+1$-dimensional $U(3)$ gauge 
theory. The D-string is a magnetic monopole of
charges $(1,-1,0)$ under the $U(1)\times U(1)\times U(1)$ Cartan
subalgebra of $U(3)$, while the \Dbar-string is a monopole of
charges $(0,-1,1)$. In terms of the $SU(3)$ subgroup of $U(3)$
(neglecting the centre-of-mass factor), the configuration is like a
monopole of one $U(1)$ and an anti-monopole of the other. With this 
description, one may calculate the force between the two D-strings 
by treating the monopoles as point particles \refs\mons. In fact, the 
result may be seen quite simply, as both D-string and \Dbar-string 
appear as positive charges on the middle D3-brane. We therefore 
expect them to repel, and this is indeed the case.

Let us now be both more quantitative and more general. Consider 
an arbitrary simple gauge group 
${\cal G}$, with a single adjoint Higgs field\foot{Of course, the 
$U(N)$ gauge theory on D3-brane has 6 adjoint Higgs fields. We assume 
that all branes are co-linear. The force between two D-strings (as 
opposed to D-string and anti-D-string) when this is not the case 
has been calculated in \refs\christim.} $\phi$, which acquires a 
vacuum expectation value (VEV), $\langle\phi\rangle ={\bf h}\cdot{\bf H}$, 
where ${\bf H}$ is the rank $r$ dimensional Cartan basis and ${\bf h}$ 
is an $r$-dimensional vector. 
We assume the VEV is such that ${\cal G}$ is broken to the
maximal torus, $U(1)^r$.

The vector ${\bf h}$ lies in the root space and therefore 
determines a fundamental Weyl chamber from which we define the simple
roots, $\alpha_j$, $j=1,\cdots,r$. Recall that these are the roots
that satisfy ${\bf h}\cdot\alpha_j >0$ and have the property
that any other root is a linear combination of the $\alpha$'s with
either all positive or all negative coefficients.

Magnetic monopoles are configurations with magnetic field given 
asymptotically by,
\eqn\asymp{
B_i={\bf g}\cdot{\bf H}{{r_i}\over{4\pi r^3}}}
The magnetic charge vector ${\bf g}$ is forced by topological 
considerations to lie in the co-root lattice, 
${\bf g}=4\pi\sum_jn_j\alpha^\star_j/e$, where $n_j$ are integers 
and $e$ is the gauge coupling constant. Whether a given topological 
charge sector is to be considered as a monopole or anti-monopole 
depends on the sign of ${\bf h}\cdot{\bf g}$, as can be seen in 
the most general form of the Bogomol'nyi equation,
\eqn\bog{
B_i=\sign({\bf h}\cdot{\bf g})\ D_i\phi}
In particular, all charges equal to linear combinations of positive co-roots 
are monopoles, while those equal to linear combinations of negative 
co-roots are anti-monopoles. We 
will be interested in charges that are given by the sum of a positive 
co-root and negative co-root. Notice that for $r>1$, there are non-trivial 
topological sectors with this property. This contrasts with the case 
of $SU(2)$ gauge group where a monopole-antimonopole pair necessarily 
lies in the sector with zero topological charge.

The method used to calculate the long-range force between two static
monopoles of magnetic charges ${\bf g}_1$ and ${\bf g}_2$ separated by
a distance $r$, is well known \refs\mons.  Treating the monopoles as
point particles, they interact through two massless fields: the gauge
field and the Higgs field.  We treat each in turn. Firstly, the
magneto-static potential is given by
\eqn\vem{
V_{em}={\bf g}_1\cdot{\bf g}_2 {{1}\over{4\pi r}} }
Secondly, the potential due to the massless scalar field is determined 
by approximating the configuration of two monopoles by a simple 
superposition of the individual solutions. Using Eqn. \asymp\ and 
\bog, the asymptotic form of the Higgs field of the second 
monopole is given by,
\eqn\phiass{
\phi={\bf h}\cdot{\bf H}-\sign ({\bf h}\cdot{\bf g}){{1}\over{4\pi r}}
{\bf g}\cdot{\bf H}}
The potential is calculated by examining the energy of the first monopole 
which, when isolated, is given by $M=\|{\bf h}\cdot{\bf g}_1\|$. 
However, in the presence of the second monopole, the effective 
mass of the first becomes,
\eqn\meff{
M_1^{eff}=\|\left({\bf h}-\sign ({\bf h}\cdot{\bf g}_2)
{{1}\over{4\pi r}}{\bf g}_2\right)\cdot{\bf g_1}\|}
which, for large separation $r\gg1/\|{\bf h}\|$, has an expansion as
\eqn\meffagain{
\eqalign{
M_1^{eff}&=\|{\bf h}\cdot{\bf g}_1\|-\sign ({\bf h}\cdot{\bf g}_1)
\,\sign ({\bf h}\cdot{\bf g}_2){{{\bf g}_1\cdot{\bf g}_2}\over{4\pi r}} \cr 
&=M_1+V_{Higgs}\cr}}
The full potential between two well separated monopoles of charges 
${\bf g}_1$ and ${\bf g}_2$ is therefore
\eqn\vfull{
V=V_{em}+V_{Higgs}={{{\bf g}_1\cdot{\bf g}_2}\over{4\pi r}}
\left(1-\sign({\bf h}\cdot{\bf g}_1)\,\sign({\bf h}\cdot{\bf g}_2)\right)}
Let us now illustrate this formula with a few simple examples. 
Firstly, consider 
an $SU(2)$ gauge theory with the VEV $h>0$. We have two possibilities: 
$g_1=g_2=1$ or $g_1=-g_2=1$. In the first situation, the potential vanishes. 
This is simply the well-known cancellation of the force between two 
BPS monopoles. In the second situation however, the magneto-static 
and Higgs potentials combine to produce a negative potential, reflecting 
the attractive force between a monopole-antimonopole pair. This is the 
relevant situation for Fig.(2.3). 

Now applying Eqn.\vfull\ to the case at hand, we choose ${\bf h}$
such that the simple roots of $SU(3)$ are $(1,-1,0)$ and $(0,1,-1)$.
The sector we are interested in consists of a monopole of charge ${\bf
g}_1=(1,-1,0)$, and another of charge ${\bf g}_2=(0,-1,1)$.  We
therefore have $\sign({\bf h}\cdot{\bf g}_1)=+1$ while $\sign ({\bf
h}\cdot{\bf g}_2)=-1$ and again we see non-cancellation between the
magneto-static and Higgs forces. However, now the potential is
positive. The adjacent D-string and anti-D-string therefore repel.

This result is interesting in itself, but we can ask what it tells us
about the original system of an adjacent \ddb4\ pair. As we observed
already, these calculations do not apply directly to the system that
we started out with, because the configuration they describe is
related to our original one only after S-duality. However, we can make
two relevant observations. One is that if we take a system of parallel
D5-branes in type IIB and suspend a D3 brane in one interval and a
$\dbar3$ in the next, then we only need T-dualities, and no
S-dualities, to relate that system to the $SU(3)$
monopole-antimonopole pair studied above. In this way we can remain at
weak coupling throughout, hence the above computation reliably tells
us that an adjacent brane-antibrane pair separated by a D5-brane does
repel.

The second observation is that even for an adjacent brane-antibrane
pair separated by an NS5-brane, the qualitative reason for repulsion
still holds: charges of the same sign are deposited on the middle
brane, while on the outer branes the product of deposited charges is
zero. Thus, just by analysing charges, one sees that the pair should
repel. We will confirm this by dualising to fractional branes in a
subsequent section.

An interesting extension of the above model(s) arises as
follows. Supposing we start with four parallel NS5-branes in type IIA
theory. Then we can suspend a D4-brane between the first two and a
$\dbar4$-brane between the next two (Fig.(4.3)). In this situation the
\ddb4\ pair is not adjacent, but separated by an ``empty''
interval. The arguments of the previous section, based on charges,
suggest that at tree level there is neither attraction nor repulsion
between the \ddb4\ pair in this model. Hence we appear to have found a
stable, non-BPS configuration from brane-antibrane constructions.
Indeed, in the related D1-D3 system, there exists a corresponding
non-BPS solution of the classical field equations, consisting of a
monopole and anti-monopole placed in commuting $SU(2)$ subgroups.
\bigskip

\centerline{\epsfbox{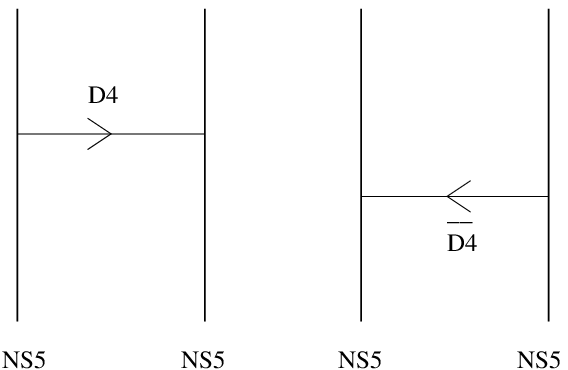}}\nobreak
\centerline{\footnotefont 
Fig.(4.3): Non-adjacent \ddb4\ pairs in type IIA.}
\bigskip
Note that in this system, the locations of the first two NS5-branes in
$x^7,x^8,x^9$ must be equal to the location of the D4-brane in these
directions, while the locations of the next two NS5-branes in these
directions must likewise be equal to that of the $\dbar4$-brane ending
on them. However, these two sets of locations need not be equal to
each other. 

This system can be shown to be T-dual to a $Z_4$ ALE
singularity. The above D4 and $\dbar4$ branes correspond to two of the
eight different fractional branes that arise when a full \ddb4\ pair
is brought to a $Z_4$ singularity. The spectrum of light states for
strings stretching between such a pair can in principle be analysed
using quiver techniques, though we will not do so in this paper.

Note that if we replace the D4 and $\dbar4$ branes with D2 and
$\dbar2$ branes respectively, we end up with a system that is similar
to one that was investigated in Ref.\refs\sencycle. In the latter, a
well-separated pair of $Z_2$ ALE singularities arranged at opposite
ends of a circle is shown to support a stable non-BPS state,
consisting of a fractional D2-brane wrapped on the first singularity
and a $\dbar2$-brane wrapped on the second. As the singularities get
closer, this system is unstable to decay into a single D-string of
type IIA wrapped on the circle that connects them. As we will see,
well-separated ALE singularities can be T-dualised to a set of
NS5-branes on a very small circle, which is precisely the system being
considered in this paper. The limit of a small circle ensures that the
approximation of a common world-volume field theory is valid. While
the system described in Fig.(4.3) is not dual to a pair of $Z_2$
singularites, for some purposes a $Z_4$ singularity with the two end
cycles shrunk to zero size should provide similar physics. It would be
interesting to explore whether a similar phase diagram to the one
studied in Ref.\refs\joysen\ arises in the $Z_4$ model.

\newsec{Borromean Branes}

Consider now a configuration in type IIB string theory containing
NS5-branes, D5-branes and D3-branes all together. The branes are
aligned as follows:
\eqn\branealign{
\eqalign{
NS5:\qquad& (x^1,x^2,x^3,x^4,x^5)\cr
D5:\qquad& (x^1,x^2,x^7,x^8,x^9)\cr
D3:\qquad& (x^1,x^2,x^6)\cr}}
This is the class of configurations studied by Hanany and
Witten\refs\hanwit, and following them we also use the notation
$\vx=(x^3,x^4,x^5)$ and $\vy=(x^7,x^8,x^9)$. Scalar fields
representing translational modes in these directions will be denoted 
$\vX$ and $\vY$. 

To start with, consider two parallel NS5-branes at the same value of
$\vy$, say $\vy=\vzero$, and separated by a finite distance in
$x^6$. Suspend a D3-brane between them. This can be located at any
value of $\vx$, though it must have $\vy=0$ to end on the NS5-branes.
At some intermediate value of $x^6$, insert a D5-brane at a fixed
$\vx$, say $\vx=\vzero$. The whole configuration is shown in Fig.(5.1).
\bigskip

\centerline{\epsfbox{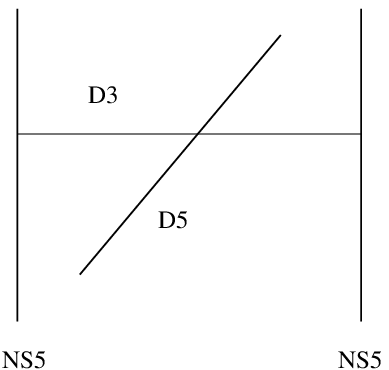}}\nobreak
\centerline{\footnotefont 
Fig.(5.1): Hanany-Witten configuration.}
\bigskip
Any two of the three branes in this construction
break the supersymmetry to $1\over 4$ of the maximal value, so the
field theory on the common intersection will have 8
supersymmetries. Adding the third brane does not further reduce the
number of supersymmetries. To illustrate this, split the 32 supercharges 
of IIB string theory into four groups of 8, labelled $Q_\lambda$, 
$\lambda =1,\cdots,4$. Each brane breaks half of the supersymmetry 
as indicated in the following table,
\eqn\fivethree{
\matrix{& & Q_1 && Q_2 && Q_3 && Q_4\cr
\noalign{\medskip}
NS5 & & \times &&\tick && \times && \tick\cr
\noalign{\smallskip}
D5 & & \tick && \times && \times && \tick\cr
\noalign{\smallskip}
D3 & & \times && \times && \tick && \tick\cr}}
where a $\times$ denotes that the supersymmetry is broken by a
given brane, while a $\tick$ means that it is preserved. We see that the 8
supercharges of $Q_4$ are preserved by all branes, and we therefore
have a $2+1$-dimensional field theory with $\cN=4$ supersymmetry
describing the low energy dynamics of the D3-brane.

The massless spectrum of this $2+1$-dimensional field theory is as
follows\refs\hanwit. Firstly, the 3-brane suspended between two
parallel NS5-branes has a pure $\cN=4$ supersymmetric $U(1)$ gauge
theory, containing a gauge field, three scalars associated to the
transverse motion of the 3-brane in the $\vx$ directions, and
fermions. Inserting the D5-brane in the middle gives rise to a charged
hypermultiplet from the open string connecting it to the
D3-brane. This hypermultiplet becomes massless when the D3-brane moves
to $\vx=0$, where it touches the D5-brane.

Now let us ask what happens if we replace the D3-brane with a
$\dbar3$-brane, while keeping the NS5 and D5 branes unchanged
(Fig.(5.2))\foot{This configuration was described in
Ref.\refs\gaunthull.}. The supersymmetries preserved by the system may
be read off from Table \fivethree, simply by exchanging each $\times$
and $\tick$ in the D3-brane row, to get
\eqn\fivethreebar{
\matrix{& & Q_1 && Q_2 && Q_3 && Q_4\cr
\noalign{\medskip}
NS5 & & \times &&\tick && \times && \tick\cr
\noalign{\smallskip}
D5 & & \tick && \times && \times && \tick\cr
\noalign{\smallskip}
\dbar 3 & & \tick && \tick && \times && \times\cr}}
The system now breaks all supersymmetries. Notice however that any
pair of branes still preserves 8 supercharges. We call this
configuration ``Borromean'' (in analogy with the famous topological
configuration of three rings, where any two are unlinked but all three
together are linked). We may restore supersymmetry by
changing either of the 5-branes to an antibrane. If however we
change both D5 and NS5-branes to $\dbar5$ and ${\bar{\rm NS}}5$-branes
respectively, so that all branes in Fig.(5.1) have been changed to
antibranes, we once again have a situation preserving no
supersymmetry. 
\bigskip

\centerline{\epsfbox{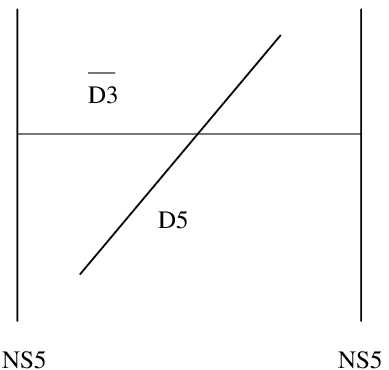}}\nobreak
\centerline{\footnotefont 
Fig.(5.2): Borromean branes.}
\bigskip
Let us now analyse the spectrum of the Borromean brane system. Since
the $\dbar3$-brane suspended between two NS5-branes is supersymmetric,
it will give rise to an $\cN=4$ $U(1)$ vector multiplet. On the other
hand, if we consider the $\dbar3$-brane along with the D5-brane, the
pair is also supersymmetric and the open string joining them gives
rise, as previously, to an $\cN=4$ hypermultiplet. Since the light
fields all come from considering two of the three branes at a time, we
see that there is no tachyon in the one-particle Hilbert space of this
theory. However, although pairwise supersymmetric, the coupling of the
hypermultiplet to the vector multiplet does not preserve
supersymmetry. We will now argue that the correct non-supersymmetric
$d=2+1$-dimensional low-energy gauge theory is given by the usual
$\cN=4$ Lagrangian {\it without} the Yukawa terms coupling a
hypermultiplet scalar and fermion to a vector multiplet fermion.

To see this we use the pairwise supersymmetry of the configuration to
determine which fermions on the worldvolume of the D3-brane couple to
the 3-5 strings, and which are projected out by the boundary
conditions of the NS5-branes. Let us begin with the latter problem and
consider an infinite 3-brane. The world-volume fields on this brane
are all goldstone modes for broken symmetries: 6 scalars, $\vX$ and
$\vY$, and 16 goldstinoes (corresponding to the supercharges $Q_1$ and
$Q_2$ for the D3-brane of Eqn.\fivethree, and to $Q_3$ and $Q_4$ for the
$\dbar3$-brane of Eqn.\fivethreebar). The presence of the NS5-branes
on which the 3-brane ends projects out half of these fields. Of the
bosons, the surviving fields are $\vX$, which are the space-time
directions which do not arise as goldstone bosons on the NS5-brane,
together with the $A_0,A_1$ and $A_2$ component of the gauge field.
By supersymmetry, the surviving fermions are those goldstinoes which
arise from supersymmetries broken by the 3-brane, but preserved by the
NS5-branes. In the case of the D3-brane, this means $Q_2$, while for
the $\dbar3$-brane, it is $Q_4$.

Now let us examine the hypermultiplet fields arising from the 3-5
string. We are interested in determining to which fields on the
3-brane they couple. Once again, we consider first the case of an
infinite 3-brane. The world-volume theory is $d=3+1$, $\cN=4$ $U(1)$
gauge theory. The presence of the D5-brane breaks this supersymmetry
by half and the $\cN=4$ multiplet splits into an $\cN=2$ vector
multiplet and a neutral $\cN=2$ hypermultiplet. The manner in which
this split occurs is determined by the fact that the charged
hypermultiplet arising from the 3-5 string is minimally coupled only
to the former. Notice that as the D5-brane is not Lorentz invariant
from the point of view of an infinite 3-brane world-volume, the gauge
field will also be split, partly living in the vector multiplet and
partly in the hypermultiplet.  More specifically, the 3-5 string
couples to the components of the gauge field $A_0,A_1$ and $A_2$,
together with the scalars $\vX$. To see the latter is particularly
simple, as motion of the 3-brane in the $\vx$ directions stretches the
3-5 string and so gives a mass to the charged hypermultiplet.  Stated
differently, the charged 3-5 hypermultiplet couples to those scalars
which are goldstone modes for both the 3-brane and the D5-brane.  By
supersymmetry, it therefore also couples to the fermions on the
3-brane which are goldstinoes for both 3-brane and D5-brane. Hence, in
the case of the D3-brane in \fivethree, the 3-5 string couples to
$Q_2$, while for the $\dbar3$-brane of
\fivethreebar, it couples to $Q_3$.

Putting these two facts together allows us to determine the field
theory for the Borromean branes of Fig.(5.2). The 3-5 string couples
to the scalars $\vX$, the gauge fields $A_0,A_1,A_2$ and the
goldstinoes $Q_3$. However, the NS5-branes project out all fields
except for the above bosons and the goldstinoes $Q_4$. In particular,
the fermions to which the hypermultiplet couples are lost due to the
boundary conditions imposed by the NS5-brane. Therefore, the gauge
theory on the D3-brane is as in the supersymmetric case, except that
the Yukawa terms coupling the 3-5 string hypermultiplet to the 3-3
string fermions is missing, thus breaking supersymmetry.

Having identified the non-supersymmetric gauge theory living on the
D3-brane, we must now determine the correct vacuum state of this
theory.  Classically, the theory has flat directions given by the
vacuum expectation values of the scalars $\vX$ and the dual photon. As
the theory is non-supersymmetric, the flat directions associated to the 
scalars $\vX$ are not protected
against quantum corrections. However, they do remain at one-loop as a
consequence of bose-fermi degeneracy. Whether this cancellation
continues at higher loops is an open question.

One may pass from the Borromean brane situation to that of adjacent
brane-antibrane pairs, generalising those discussed in the previous
section, by passing the D5-brane through one of the NS5-branes. The
pairwise supersymmetry of these two 5-branes ensures that the
Hanany-Witten transition\refs\hanwit\ occurs as in the supersymmetric
case, and we are left with a D3-brane suspended between the D5 and
NS5-brane (Fig.(5.3)). 
\bigskip

\centerline{\epsfbox{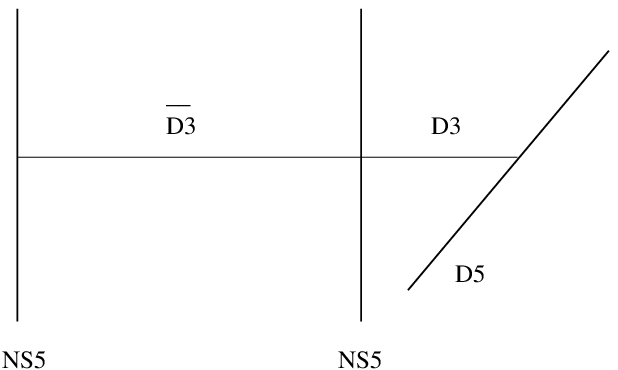}}\nobreak
\centerline{\footnotefont 
Fig.(5.3): Result of Hanany-Witten transition on Borromean branes.}
\bigskip
The brane configuration is no longer Borromean, as the D3 and
$\dbar3$-branes are not pairwise supersymmetric. Applying the
considerations of the previous section locally to the region near the
middle 5-brane, we would expect that the \ddb3\ pair repels.  However,
it is quite possible that in this model the Hanany-Witten transition
is accompanied by a phase transition, so we are unable to relate the
physics of Figs.(5.2) and (5.3). Note in particular that the
dimensionless expansion parameter in the theory on the Borromean
branes is $e^2/r$, where $e$ is the gauge coupling constant, and $r$
the distance between the D5-brane and $\dbar3$-brane.  The one-loop
flatness of the vacuum moduli space of this theory ensures that any
potential that is generated at large distance is of the form $\pm
1/r^n$, for $n\geq 2$. This is to be contrasted with the $1/r$
behaviour of adjacent brane-antibrane pairs separated by a D5-brane.

\newsec{Dualising NS5-branes into ALE spaces}

Consider type IIA string theory compactified on a 4-torus $T^4$ in the
directions $x^6,x^7,x^8,x^9$ and orbifolded by the inversion
$I_{6789}$. This gives rise to 16 fixed orbifold 5-planes. Each of
these is locally a $Z_2$ ALE space. Or goal is to dualise this system
and then, by going to a suitable region of moduli space, to isolate
the dual of a single $Z_2$ ALE space. The material in this section is
not new, having been discussed in slightly different language in
Refs.\refs\nsdual\ and in the context of conifold singularities in
Refs.\refs{\urangaconif,\dmconif}. The duality that we discuss below
will be used in the subsequent section to map the suspended brane
systems to brane-antibrane generalisations of quiver theories.

Under T-duality along $x^6$, the inversion $I_{6789}$ is mapped to
$\mofl I_{6789}$, as one can easily see by examining the action on
various massless states. At the same time, the type IIA theory is
mapped to the type IIB theory. Thus the dual to the original orbifold
is the orbifold of type IIB string theory on a 4-torus quotiented by
$\mofl I_{6789}$. This too has 16 fixed points, but the nature of the
orbifold 5-planes is quite different. To see this, perform an
S-duality. The operator $\mofl$ is mapped to the orientifolding
operation $\Omega$, and the resulting theory has 16 orientifold
5-planes (O5-planes) along with 16 (mirror pairs of) D5-branes. The
D5-branes can sit on the orientifold planes to make charge-cancelled
configurations, or they can move off. Hence it must be that, before
performing S-duality, the 16 orbifold planes were made up of
NS5-branes (which are S-dual to D5-branes in type IIB) and static
planes that are S-dual to O5-planes. We call these ``S5-planes''.

To summarise, the T-dual of type IIA on a $T^4/Z_2$ orbifold is type
IIB on a 4-torus with 16 S5-planes and 16 pairs of NS5-branes. At the
orbifold point of the type IIA theory, the flux of the 2-form
$B$-field through each $Z_2$ singularity is $\half$, while on the IIB
side, each NS5-brane is on top of an S5-plane. Now since the T-duality
was performed only on $x^6$, we can identify the $x^7,x^8,x^9$
positions on the two sides and decompactify them. As a result we have
on the type IIA side, a pair of $Z_2$ ALE spaces located symmetrically
around the $x^6$ circle, and at the origin of $x^7,x^8,x^9$, while on
the type IIB side there is a pair of S5-planes similarly located
symmetrically around the dual $x^6$ circle, and a pair of NS5-branes
at points of this circle (Fig.(6.1)). 
\bigskip

\centerline{\epsfbox{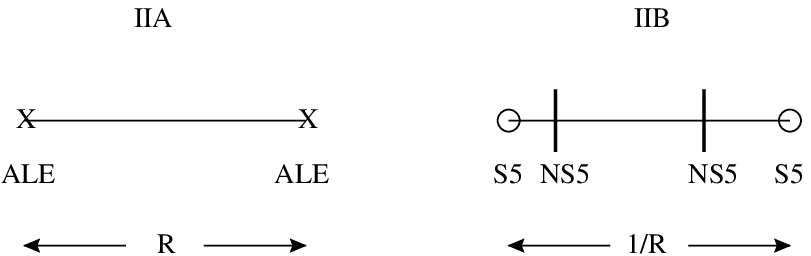}}\nobreak
\centerline{\footnotefont 
Fig.(6.1): Duality between $Z_2$ ALE spaces and 
S5-NS5 systems.}
\bigskip
The matching of moduli is reasonably straightforward. In particular,
if $R$ is the radius of the $x^6$ circle on the type IIA side then the
dual radius is ${1\over R}$. Let $B,B'$ be the fluxes of the
$B$-field through the two $Z_2$ ALE spaces on the IIA side, and let
$X_6^{(1)}, X_6^{(2)}$ be the $x_6$ locations of the two NS5-branes on 
the IIB side. Then we have:
\eqn\modmatch{
\eqalign{
X_6^{(1)} &= {1\over R} \vert B-B'\vert\cr
X_6^{(2)} &= {1\over R} \vert B+B'\vert\cr}}
Note that with this matching, the type IIA theory has an enhanced
gauge symmetry if $B=0$ or $B'=0$, coming from a BPS D2-brane wrapping
the 2-cycle of the first or second ALE space, and this corresponds in
type IIB to $X_6^{(1)}= X_6^{(2)}$ so the two NS5-branes meet. If both
$B=B'=0$ then we have $SU(2)\times SU(2)$ enhanced gauge symmetry,
and on the type IIB side, $X_6^{(1)}= X_6^{(2)}=0$, so the two
NS5-branes meet on an S5-plane, where in fact we expect $SO(4)\sim
SU(2)\times SU(2)$.

Suppose we now fix $B'=\half$ once and for all. Then no enhanced gauge
symmetry will ever arise from the second ALE space, and henceforth we
focus our attention on the region near the first ALE space. The dual
picture now has the two NS5-branes at equal distances from their
corresponding S5-planes. They can only meet far away from the
S5-planes (in fact, at the midpoint of the strip depicted in
Fig.(6.1)). Hence the S5 orbifold planes can be ignored. The result is 
a duality between a pair of NS5-branes arranged around a circle (in
type IIB) and a solitary $Z_2$ ALE singularity (in type IIA). The
roles of type IIA and type IIB can also be reversed. 

Now we may wrap a \Dp-brane around the $x^6$ circle and perform the
above duality. It turns into a D$(p-1)$ brane at the $Z_2$ ALE
singularity. Breaking the original $p$-brane on the NS5-branes
corresponds to breaking the $(p-1)$-brane into a pair of fractional
branes, which are free to move around while remaining in the 5-plane
of the ALE singularity. 

In a similar fashion, one can start with a 2-torus slanted at a
particular angle and carry out a $Z_k$ quotient $z\rightarrow \omega
z$ where $\omega^k=1$. Near the origin, the system can be described in
T-dual language by a set of $k$ NS5-branes located around the $x^6$
circle. The separations of the NS5-branes along $x^6$ are given by
B-fields through the various 2-cycles of the $Z_k$ singularity, while
their locations in $x^7,x^8,x^9$ are determined by the geometrical
size parameters of these 2-cycles.
 
\newsec{Brane-Antibrane Pairs at an ALE Singularity}

In this section we obtain the spectrum of light fields (including
possible tachyons) on brane-antibrane pairs placed at ALE quotient
singularites. For simplicity we will work mainly with the case of
$R^4/Z_2$. We start with this singularity, where $R^4$ corresponds to
the directions $x^6,x^7,x^8,x^9$, and bring a \ddb{p}\ pair close to
it ($p$ is even in type IIA and odd in type IIB). Our analysis will
closely parallel that of Ref.\refs\dougmoore\ where BPS D-branes at
ALE spaces were studied. Some aspects of brane-antibrane systems at
ALE spaces have been investigated previously in Ref.\refs\aldauranga.

In order to work out the spectrum of the resulting theory, we first
introduce the ``mirror'' of the \ddb{p}\ pair in the $Z_2$
singularity, thus making four branes altogether. Let these four branes
be labelled $1,1',\onebar,\onebar'$, where the prime denotes a mirror
brane and the bar denotes an antibrane. Thus there are $4\times 4$
Chan-Paton factors, which we classify as follows. We use the $2\times
2$ Chan-Paton labels $A,B,C,D$ to distinguish between strings on a
brane, on a mirror brane, or between the two, while the labels
$a,b,c,d$ are used (as before) to distinguish between strings on a
brane, on an antibrane, or between the two.  The correspondence
between $a,b,c,d$ and Chan-Paton matrices was given in Eqn.\cpfactors,
and the analogous correspondence holds also for $A,B,C,D$.

With these definitions, we have the correspondence:
\eqn\hbcpfactors{
\matrix{1-1: A\otimes a\cr 1-1': C\otimes a\cr
1'-1: D\otimes a\cr 1'-1': B\otimes a\cr}
\quad\quad
\matrix{\onebar-\onebar: A\otimes b\cr
\onebar-\onebar': C\otimes b\cr
\onebar'-\onebar: D\otimes b\cr
\onebar'-\onebar': B\otimes b\cr}
\quad\quad
\matrix{1-\onebar:A\otimes c\cr 1-\onebar': C\otimes c\cr
1'-\onebar: D\otimes c\cr 1'-\onebar': B\otimes c\cr}
\quad\quad
\matrix{\onebar-1: A\otimes d\cr
\onebar-1': C\otimes d\cr
\onebar'-1: D\otimes d\cr
\onebar'-1': B\otimes d\cr}}

The spectrum of this system, before projection by the $Z_2$, contains
gauge fields $A_\mu^{(I)}$, $\mu=0,\ldots,p$, along with scalars
$X_i^{(I)}$, $i=p+1,\ldots,5$, and $X_m^{(I)}$, $m=6,7,8,9$. These
fields all lie in the adjoint of $U(2)\times U(2)$, with the
superscript $(I)$ labelling the relevant factor. Along with the
obvious fermionic counterparts, these make up a set of massless fields
arising in the Chan-Paton sectors that appear in the first two columns
of Eqn.\hbcpfactors. Each Chan-Paton sector in the remaining two
columns gives rise to a tachyon $T$, hence there are 8 real tachyons
altogether. These are accompanied by a set of massless Ramond fermions
which are oppositely GSO-projected as compared to the ones that
accompanied the gauge field and massless scalars.

There are two interesting $Z_2$ involutions: the inversion that
creates the ALE space, which we denote by $I_{6789}$, and the symmetry
$\mofl$ which we have used before. In our conventions, these
involutions act as conjugation by:
\eqn\invol{
\eqalign{
I_{6789}&: \sigma_1\otimes 1\cr
\mofl &: 1\otimes \sigma_1\cr}}
This follows from the fact that the first $\sigma_1$ factor exchanges
a brane with its mirror, while the second $\sigma_1$ exhanges a brane
with an antibrane.

It is now straightforward to write down the transformations of the
various fields under $I_{6789}$:
\eqn\vartransf{
\matrix{\underline{\rm Field}\hfill & &\underline{\rm CP~factor}\hfill & & 
\underline{I_{6789}}\hfill\cr
&&&\cr
A_\mu^{(I)}, X_i^{(I)}\hfill & &\{1,\sigma_1\}\otimes\{1,\sigma_3\}\hfill
& & +\hfill\cr
X_m^{(I)}\hfill & &\{i\sigma_2,\sigma_3\}\otimes \{1,\sigma_3\}\hfill &
& +\hfill\cr 
A_\mu^{(I)}, X_i^{(I)}\hfill & &\{i\sigma_2,\sigma_3\}
\otimes\{1,\sigma_3\}\hfill & & -\hfill\cr
X_m^{(I)}\hfill & &\{1,\sigma_1\} \otimes \{1,\sigma_3\}\hfill & &
-\hfill\cr 
T\hfill & &\{1,\sigma_1\}\otimes \{\sigma_1,i\sigma_2\}\hfill & &
+\hfill\cr 
T\hfill & &\{i\sigma_2,\sigma_3\}\otimes \{\sigma_1,i\sigma_2\}\hfill &
& -\hfill\cr}}

Thus under orbifolding by $I_{6789}$, the fields which are invariant
survive, while the others are projected out. The gauge group that
survives (for a brane-antibrane pair on top of the orbifold
singularity) is $U(1)^4$. This statement really applies when all vev's
are zero, which is the common origin of the Higgs branch and Coulomb
branch. We see from the table that 4 real tachyons survive in the
orbifolded theory, these can be grouped into two complex tachyons that
transform as the bi-fundamental of the first $U(1)\times U(1)$ and the
second $U(1)\times U(1)$ respectively.

On the Higgs branch, the brane-antibrane pair leaves the orbifold
plane and the gauge group is Higgsed to $U(1)\times U(1)$. This
corresponds to retaining the following fields from the above table:
\eqn\higgsbr{
\eqalign{
A_\mu^{(I)}, X_i^{(I)}&:\quad 1\otimes\{1,\sigma_3\} \cr
X_m^{(I)}&:\quad \sigma_3\otimes \{1,\sigma_3\} \cr
T&:\quad 1\otimes \{\sigma_1,i\sigma_2\}\cr}}
(For $X_m$ this involves a gauge choice, see Ref.\refs\polkthree.)
Thus on this branch there is a single complex tachyon, transforming as
the bi-fundamental of $U(1)\times U(1)$.

We can relate all this to the brane constructions that we discussed
above. Compactify $x^6$ on a circle and T-dualize, then we have a pair
of NS5-branes on the circle and a \ddb{(p+1)}\ pair running all the way
around the circle (Fig.(6.1)). 
\bigskip

\centerline{\epsfbox{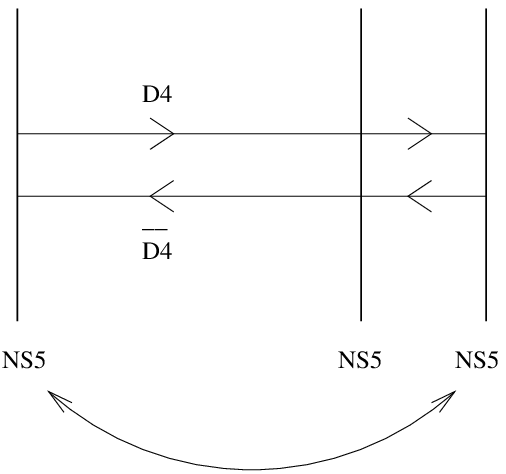}}\nobreak
\centerline{\footnotefont 
Fig.(7.1): A pair of NS5-branes and a \ddb{(p+1)}\ pair wrapped 
on $x^6$.}
\bigskip
The figure describes the point where the Higgs branch and Coulomb
branch meet, and it is clear that the gauge group should be $U(1)^4$
as we have shown. Moreover, in this brane construction one expects 2
complex tachyons, corresponding to open strings connecting a brane and
antibrane segment suspended between the same NS5-branes. They
manifestly carry the right charges. Moving onto the Higgs branch is
accomplished by taking the brane-antibrane pair away from the
NS5-branes, this breaks the gauge group to $U(1)\times U(1)$ and there
is just one complex tachyon.

To understand the Coulomb branch, observe that both the brane and the
antibrane which wrap all the way around $x^6$ in the dual brane
construction can break on the NS5-branes. (In the T-dual picture this
corresponds to the fact that branes at an ALE singularity can break
into fractional
branes\refs{\polkthree,\dougmoore,\fracbranes,\dmfrac}. Hence, altogether we
have four brane segments in the problem (or rather, two brane segments
and two antibrane segments). We can separate all these and take some
of the segments off to infinity. For example, suppose we take a brane
and an antibrane stretching along the same segment to infinity. The
result (after taking the radius very large) will be the diagram of
Fig.(2.1). On the other hand, we can take away an antibrane from one
segment and a brane from the adjacent segment. The result will be an
adjacent brane-antibrane pair, similar to the construction in
Fig.(4.1). We will use these facts in the T-dual picture to understand
more about the brane constructions of Figs.(2.1),(4.1).

Returning to the \ddb{p}\ pair at an ALE singularity, we should now
identify the Chan-Paton factors corresponding to strings connecting
fractional branes. Recall that for a single \Dp-brane at a $Z_2$
singularity, the Coulomb branch arises when $X_m=0$ and $X_i
\sim \{1,\sigma_1\}$. It is on this branch that the single \Dp-brane is
said to split into a pair of fractional branes and the gauge group
becomes $U(1)\times U(1)$. Note that the individual gauge fields
associated to the two fractional branes are associated to the
Chan-Paton factors $1+\sigma_1$ and $1-\sigma_1$ respectively.

For a \ddb{p}\ system, the Higgs branch is described by
$X^{(I)}_m\ne0$, $X^{(I)}_i\sim 1\otimes\{1,\sigma_3\}$ while the
Coulomb branch is $X^{(I)}_m=0$, $X^{(I)}_i \sim \{1,\sigma_1\}\otimes
\{1,\sigma_3\}$. This time, as expected, the latter branch describes
four fractional branes altogether. Let us denote these fractional
branes as $1_f,1'_f,\onebar_f$ and $\onebar'_f$ (note that in this
case, the prime does {\it not} denote the mirror image, though the bar
still denotes the antibrane). The interpretations of these branes are
as follows: $1_f$ is a (p+2)-brane wrapped on the vanishing 2-cycle
$\Sigma$ at the ALE singularity\foot{At the orbifold point, we have an
unresolved orbifold singularity through which the $B$-field has a flux
of $\half$.  This also means that the NS5-branes are symmetrically
placed along the $x^6$ circle in the T-dual construction. However, we
can of course allow $B$ to vary, and in the following discussion we
have in mind this more general case.}. $1'_f$ is an anti-(p+2)-brane
wrapped on the same 2-cycle. On the other hand, $\onebar_f$ is an
anti-(p+2)-brane, while $\onebar'_f$ is a (p+2)-brane.

Following the discussion in Ref.\refs\dmfrac, we can assign the
following world-volume couplings to the four fractional branes:
\eqn\worldcoup{
\eqalign{
&1_f:\qquad \int\,\left(\crr{p+3} + (B_{NS}-F_{1_f})\right)\wedge
\crr{p+1} \cr
&1'_f:\qquad \int\,\left(-\crr{p+3} - (B_{NS}-F_{1'_f})\right)\wedge
\crr{p+1} \cr
&\onebar_f:\qquad \int\,\left(-\crr{p+3} -
(B_{NS}-F_{\onebar_f})\right)
\wedge\crr{p+1} \cr
&\onebar'_f:\qquad \int\,\left(\crr{p+3} +
(B_{NS}-F_{\onebar'_f})\right)
\wedge\crr{p+1} \cr}}
Moreover, the world-volume field strengths are given by
$\int_\Sigma\, F_{1_f} =\int_\Sigma\, F_{\onebar_f} = 0$ and
$\int_\Sigma\, F_{1'_f} =\int_\Sigma\, F_{\onebar'_f} = 1$.

We would like to find the spectrum arising from all possible
configurations of open strings on this set of four fractional
branes. This can be read off directly from \vartransf.  First consider
open strings starting and ending on the same brane. These each carry a
gauge field $A_\mu$ and a pair of scalars $X_i$ $(i=4,5)$ having
Chan-Paton factors:
\eqn\samebrane{
\eqalign{
1_f - 1_f &:\quad \half(1+\sigma_1) \otimes \half(1+\sigma_3)\cr
1'_f - 1'_f &:\quad \half(1-\sigma_1) \otimes \half(1+\sigma_3)\cr
\onebar_f - \onebar_f &:\quad \half(1+\sigma_1) \otimes 
\half(1-\sigma_3)\cr
\onebar'_f - \onebar'_f &:\quad \half(1-\sigma_1) \otimes
\half(1-\sigma_3)\cr} }
Next, we have in principle 6 oriented strings going from each one of
the four fractional branes to another, and each string has two
orientations, for a total of 12. However, some of these are projected
out. In fact, we get massless hypermultiplet scalars $X_m$ as follows:
\eqn\branepair{
\eqalign{
1_f - 1'_f &:\quad \half(\sigma_3 + i\sigma_2)\otimes \half
(1+\sigma_3)\cr
1'_f - 1_f &:\quad \half(\sigma_3 - i\sigma_2)\otimes \half
(1+\sigma_3)\cr
\onebar_f - \onebar'_f &:\quad \half(\sigma_3 + i\sigma_2)\otimes \half
(1-\sigma_3)\cr
\onebar'_f - \onebar_f &:\quad \half(\sigma_3 - i\sigma_2)\otimes \half
(1-\sigma_3)\cr }}
Tachyons arise from the following open strings:
\eqn\tachstrings{
\eqalign{
1_f - \onebar_f &:\quad \half(1+\sigma_1)\otimes \half
(\sigma_1+i\sigma_2)\cr
\onebar_f - 1_f &:\quad \half(1+\sigma_1)\otimes \half
(\sigma_1-i\sigma_2)\cr
1'_f - \onebar'_f &:\quad \half(1-\sigma_1)\otimes \half
(\sigma_1+i\sigma_2)\cr
\onebar'_f - 1'_f &:\quad \half(1-\sigma_1)\otimes \half
(\sigma_1-i\sigma_2)\cr}}
Thus there are four real tachyons in the Coulomb branch. The remaining
open strings correspond in principle to the following Chan-Paton
factors:
\eqn\projected{
\eqalign{
1_f-\onebar'_f &:\quad \half(\sigma_3 + i\sigma_2)\otimes 
(\sigma_1+i\sigma_2)\cr
\onebar'_f - 1_f &:\quad \half(\sigma_3 - i\sigma_2)\otimes 
(\sigma_1-i\sigma_2)\cr
1'_f-\onebar_f &:\quad \half(\sigma_3 - i\sigma_2)\otimes 
(\sigma_1+i\sigma_2)\cr
\onebar_f - 1'_f &:\quad \half(\sigma_3 + i\sigma_2)\otimes 
(\sigma_1-i\sigma_2)\cr}}
However, as one can see from \vartransf, these strings are projected out: 
the tachyon is removed by $I_{6789}$ and the massless states are
removed by the anti-GSO projection. Hence we find the novel result
that strings connecting $1_f$ to $\onebar'_f$ have no tachyonic or
massless bosonic states. 

Now let us consider first the system of a suspended \ddb{(p+1)}\ pair
as in Fig.(2.1). This is described in the T-dual version by keeping the
fractional branes $1_f$ and $\onebar_f$. We see that for all values of
$\int_\Sigma\, B_{NS}$, the total (p+2)-brane and p-brane charges add
up to zero.  The analysis above tells us that the spectrum on this
pair consists of a $U(1)\times U(1)$ gauge field, massless scalars
$X_i$ and a complex tachyon $T$. All this is consistent with the fact
that we expect the original pair to be able to annihilate
completely. Thus we confirm the heuristic picture of this system
developed in Section 2.

Turning next to the system of an adjacent \ddb{(p+1)}\ pair as in
Fig.(4.1), we see that in this case the dual description is obtained
by keeping the fractional branes $1_f$ and $\onebar'_f$ (or
equivalently $\onebar_f$ and $1'_f$, with some sign changes in the
following). In this case the picture is very different. The system has
a net (p+2)-brane charge of $+2$, and a net p-brane charge of
$2B_{NS}-1$. The latter vanishes only at the symmetric point
$B_{NS}=\half$, for which the NS5-branes in the original construction
are equally spaced around the circle. As far as the spectrum is
concerned, there are neither any tachyons nor any massless scalars
coming from open strings between the pair.

Finally, one can consider non-adjacent brane-antibrane pairs from this 
point of view by taking a brane-antibrane pair to a $Z_4$ ALE
singularity, where it can split into a total of 8 fractional
branes. We will not describe this explicitly here. 

\newsec{Boundary State Computation of Brane-Antibrane Forces}

In this section our aim is to construct the boundary states
corresponding to the fractional D3-branes and use them to compute
forces between pairs of fractional branes of the various types
discussed in the previous section. We will use the conventions of
Ref.\refs\senbound. 

For this we have to construct consistent boundary states which
describe D3-branes in the $Z_{2}$ orbifold of type IIB string theory
where the orbifolding group $Z_2$ is generated by $I_{6789}$ (which we
refer to as $R$). Boundary states at orbifolds have been classified and
constructed in great generality in Ref.\refs\gabstef. Here we will
only describe the formulae relevant for our purpose.

Let us consider a D3-brane in type IIB string theory on
$R^{9,1}$. Take the D3-brane to be along the $x^{3},x^{4}, x^{5}$
directions, and situated at the origin in $x^{6},x^{7},x^{8},x^{9}$
directions. We make a double Wick rotation $x^{0}\rightarrow ix^0$,
$x^1\rightarrow ix^1$, so that the light-cone directions are $x^1\pm
x^2$ and the D3-brane world-volume directions are all space-like. We
first define the basic boundary states in terms of which the D3-brane 
state will be defined.

In the untwisted sector these states are:
\eqn\bstate{\eqalign{
\vert k,\eta \rangle_{{NSNS;U}\atop{RR;U}} & = \exp \Bigg(
\sum_{n=1}^{\infty}
\left[-{1\over n}\sum_{\mu=0,3,4,5}
\alpha_{-n}^{\mu}\tilde{\alpha}_{-n}^{\mu} 
+ {1\over n}\sum_{\mu=6}^{9}\alpha_{-n}^{\mu}
\tilde{\alpha}_{-n}^{\mu}\right]\cr
& +i\eta\sum_{r>0}\left[-\sum_{\mu=0,3,4,5}\psi_{-r}^{\mu}
{\tilde{\psi}}_{-r}^{\mu}
+\sum_{\mu=6}^{9}\psi_{-r}^{\mu}\tilde{\psi}_{-r}^{\mu}\right] \Bigg)
\vert k,\eta \rangle_{{NSNS;U}\atop{RR;U}}^{(0)}}}
where $k \equiv (k^1,k^2,k^6,k^7,k^8,k^9)$ and $\eta=\pm$. In the NSNS
sector, the labels $n$, $r$ are
\eqn\labels{\eqalign{
n &\in Z_{+} \quad \quad \quad \hbox{for} \quad \mu =0,3,4,\ldots,9 \cr
r &\in Z_{+} -{1 \over 2} \quad \hbox{for} \quad \mu = 0,3,4,\ldots,9\cr}}
while in the RR sector they are:
\eqn\labelrr{
n, r \in Z_{+} \quad \quad \quad \hbox{for}\quad  \mu =0,3,4,\ldots,9}
The NSNS vacuum state is independent of $\eta$, while the RR vacuum 
states are defined as:
\eqn\vacdef{\eqalign{
\psi_{-}^{\mu}\vert k, \eta \rangle_{RR;U}^{(0)} &= 0 \qquad \hbox{for} 
\quad \mu = 0,3,4,5.\cr
\psi_{+}^{\mu}\vert k, \eta \rangle_{RR;U}^{(0)} &= 0 \qquad \hbox{for} 
\quad \mu = 6,7,8,9. \cr
\vert k, +\rangle_{RR;U}^{(0)} &= \prod_{\mu = 6}^{9} \psi_{-}^{\mu}\prod
_{\mu = 0,3}^{5}\psi_{+}^{\mu}\,\vert k, - \rangle_{RR;U}^{(0)}}}

Next let us turn to the twisted sector of the $Z_2$ orbifold. Here the
relevant boundary states are given by:
\eqn\tbndst{\eqalign{
\vert k,\eta \rangle_{{NSNS;T}\atop{RR;,T}} & = \exp \Bigg( 
\sum_{n>0}^{\infty}\left[ 
-{1\over n}\sum_{\mu=0,3,4,5}\alpha_{-n}^{\mu}\tilde{\alpha}_{-n}^{\mu} 
+ {1\over n}\sum_{\mu=6}^{9}\alpha_{-n}^{\mu}
\tilde{\alpha}_{-n}^{\mu}\right]\cr
& +i\eta\sum_{r>0}\left[-\sum_{\mu=0,3,4,5}
\psi_{-r}^{\mu}{\tilde{\psi}}_{-r}^{\mu}
+\sum_{\mu=6}^{9}\psi_{-r}^{\mu}\tilde{\psi}_{-r}^{\mu}\right]\Bigg)
\vert k,\eta \rangle_{{NSNS;T}\atop{RR;T}}^{(0)}}}
where $k \equiv (k^1,k^2)$. In the twisted NSNS sector, the labels
$n$, $r$ are
\eqn\labels{\eqalign{
n &\in Z_{+} \quad \quad \quad \hbox{for} \quad \mu =0,3,4,5, \cr
  &\in Z_{+} -{1 \over 2} \quad \hbox{for} \quad \mu = 6,7,8,9, \cr
r &\in Z_{+} -{1 \over 2} \quad \hbox{for} \quad \mu = 0,3,4,5, \cr
  &\in Z_{+} \quad \quad \quad \hbox{for} \quad \mu = 6,7,8,9.}}
while in the twisted RR sector they are:
\eqn\labelrr{\eqalign{
n, r &\in Z_{+} \quad \quad \quad \hbox{for}\quad  \mu =0,3,4,5, \cr
     &\in Z_{+} -{1 \over 2} \quad \hbox{for}\quad  \mu = 6,7,8,9.}}
Since in the twisted sector both NSNS and RR sectors have zero modes
we have to define the vacua carefully (as we did for the RR sector of
the untwisted sector). Define
\eqn\tnvacua{\eqalign{
\psi_{+}^{\mu}\,\vert k, - \rangle_{NSNS;T}^{(0)} &= 0 \quad \hbox{for}
\quad \mu = 6,7,8,9. \cr
\vert k, + \rangle_{NSNS;T}^{(0)} &= \prod_{\mu =6}^{9} \psi_{-}^{\mu}
\,\vert k, - \rangle_{NSNS;T}^{(0)} }}
Similarly for the RR vacuum state, define
\eqn\trvacua{\eqalign{
\psi_{-}^{\mu}\,\vert k, - \rangle_{RR;T}^{(0)} &= 0 \quad \hbox{for}
\quad \mu = 0,3,4,5,. \cr
\vert k, + \rangle_{RR;T}^{(0)} &= \prod_{\mu =0,3}^{5} \psi_{+}^{\mu}
\,\vert k, - \rangle_{RR;T}^{(0)} }}

Now we integrate these states over momentum, to define states
corresponding to branes at fixed positions. In the untwisted sector we
get: 
\eqn\etansdef{\eqalign{
\vert \eta \rangle_{NSNS;U} &= {\cal N}\int \prod_{\mu = 1,2,6}^{9}
d k^{\mu}\, {\vert k,\eta \rangle_{NSNS;U}} \cr
\vert \eta \rangle_{RR;U} &= 4i {\cal N} \int \prod_{\mu = 1,2,6}^{9}
dk^{\mu}\, \vert k, \eta \rangle_{RR;U}}}
while the corresponding states in the twisted sector are:
\eqn\etansdef{\eqalign{
\vert \eta \rangle_{NSNS;T} &= 2{\cal \tilde{N}}\int \prod_{\mu = 1}^{2}
d k^{\mu} {\vert k,\eta \rangle_{NSNS;T}} \cr
\vert \eta \rangle_{RR} &= 2i {\cal \tilde{N}} \int \prod_{\mu = 1}^{2}
dk^{\mu} \vert k, \eta \rangle_{RR;T}}}

Next we combine the above states into the appropriate GSO-invariant
linear combinations to describe D3 and $\dbar3$ branes at the ALE
space. In the untwisted sector this gives the states:
\eqn\unsrdef{\eqalign{
\vert U \rangle_{NSNS} &= {1 \over \sqrt{2}}\Big( \vert + \rangle_{NSNS;U} 
- \vert - \rangle_{NSNS;U}\Big), \cr
\vert U \rangle_{RR} &= {1 \over \sqrt{2}}\Big( \vert + \rangle_{RR;U}
+ \vert - \rangle_{RR;U}\Big), \cr}}
while in the twisted sector the GSO-invariant combinations are:
\eqn\tbndsts{\eqalign{
\vert T \rangle_{NSNS} &= {1 \over{\sqrt{2}}}\Big( \vert + \rangle_{NSNS;T}
+\vert - \rangle_{NSNS;T}\Big) \cr
\vert T \rangle_{RR} &= {1 \over{\sqrt{2}}}\Big( \vert + \rangle_{RR;T}
+\vert - \rangle_{RR;T}\Big) }}

Finally we can combine the untwisted and twisted sector boundary
states to produce states that describe branes in the full
theory. We find four independent consistent boundary states for D3,
$\dbar3$, which can be identified with the four fractional branes
$1_f, 1'_f, {\bar 1}_f, {\bar 1}'_f$. The states along with their
identifications are as follows:
\eqn\dthrees{\eqalign{
\vert {\rm D}3,+ \rangle &= {1 \over 2}\Big( \vert U \rangle_{NSNS} 
+ \vert U \rangle_{RR} + \vert T \rangle_{NSNS} + 
\vert T \rangle_{RR}\Big) \quad \quad : 1_f \cr
\vert {\rm D}3, - \rangle &= {1 \over 2}\Big( 
\vert U \rangle_{NSNS} + \vert U \rangle_{RR} 
- \vert T \rangle_{NSNS} - \vert T \rangle_{RR}\Big) 
\quad \quad : 1'_f \cr
\vert \dbar3, + \rangle &= {1 \over 2}\Big( \vert U \rangle_{NSNS} 
- \vert U \rangle_{RR} - \vert T \rangle_{NSNS} 
+ \vert T \rangle_{RR}\Big) \quad \quad :
{\bar{1}}'_f \cr \vert \dbar3, - \rangle &= {1 \over 2}\Big( 
\vert U \rangle_{NSNS} - \vert U \rangle_{RR} + 
\vert T \rangle_{NSNS} - \vert T \rangle_{RR}\Big) \quad \quad : 
{\bar{1}}_f }} 

Now we can explicitly write down the closed string tree amplitudes
between the above boundary states in terms of open string one loop
amplitudes. First, we choose $32 {\cal N}^2 = {v^{(4)}\over(2\pi)^4}$,
where $v^{(4)}$ is the infinite volume of the brane along the $x^6,
x^7, x^8, x^9$ directions. This normalization is determined by the
requirement that amplitudes between boundary states can be interpreted
as open-string traces. Next, define $\tq=\exp(-\pi t)$ and
\eqn\thetafns{\eqalign{
f_1(q) &= q^{1\over 12}\prod_{n=1}^{\infty}(1 - q^{2n})
= \eta(q^2)\cr
f_2(q) &= \sqrt {2} q^{1\over12}\prod_{n=1}^{\infty}(1 + q^{2n})
= \sqrt{\theta_2(q^2)\over \eta(q^2)}\cr
f_3(q)&= q^{-{1\over24}}\prod_{n=1}^{\infty}(1 + q^{2n+1})
= \sqrt{\theta_3(q^2)\over \eta(q^2)}\cr
f_4(q) &= q^{-{1\over24}}\prod_{n=1}^{\infty}(1 - q^{2n+1})
= \sqrt{\theta_4(q^2)\over \eta(q^2)}\cr }}
Then we find:
\eqn\fracone{\eqalign{ 
\int_{0}^{\infty} dl\, \langle {\rm D}3, + \vert e^{-lH_c}\vert {\rm
D}3, + \rangle
=& \int_{0}^{\infty} {dt \over 2t}\,\tr_{NS-R}\left({1+(-1)^{F} \over 2}\,
{{1+R}\over 2}\,{e^{-2tH_{0}}}\right) 
\cr
=& {v^{(4)}\over 32 (2\pi)^4}\int_0^\infty 
{dt\over t^3}\, \Bigg\{
{f_3(\tq)^8 - f_4(\tq)^8 - f_2(\tq)^8
\over f_1(\tq)^8}
\cr
&+ 4{f_4(\tq)^4 f_3(\tq)^4 - f_4(\tq)^4 f_3(\tq)^4
\over f_1(\tq)^4 f_2(\tq)^4}\Bigg\} }}
The amplitude between any other fractional brane and itself is given
by the same expression. Using the abstruse identity $f_3(\tq)^8 -
f_4(\tq)^8 - f_2(\tq)^8=0$, we see that this amplitude vanishes, as
expected.

Next, consider the amplitude:
\eqn\fractwo{\eqalign{
\int_{0}^{\infty} dl\, \langle {\rm D}3, + \vert e^{-lH_c}\vert 
{\rm D}3, - \rangle
=& \int_{0}^{\infty} {dt \over 2t}\,\tr_{NS-R}\left({1+(-1)^{F} \over 2}\,
{{1-R}\over 2}\,{e^{-2tH_{0}}}\right) 
\cr
=& {v^{(4)}\over 32 (2\pi)^4}\int_0^\infty 
{dt\over t^3}\, \Bigg\{
{f_3(\tq)^8 - f_4(\tq)^8 - f_2(\tq)^8
\over f_1(\tq)^8}
\cr
&- 4{f_4(\tq)^4 f_3(\tq)^4 - f_4(\tq)^4 f_3(\tq)^4
\over f_1(\tq)^4 f_2(\tq)^4}\Bigg\} }}
This gives the force between $1_f$ and $1'_f$, and is also equal to
zero. 

The next amplitude of interest is:
\eqn\fracthree{\eqalign{
\int_{0}^{\infty} dl\, \langle {\rm D}3, + \vert e^{-lH_c}\vert 
\dbar3, - \rangle
=& \int_{0}^{\infty} {dt \over 2t}\,\tr_{NS-R}\left({1-(-1)^{F} 
\over 2}\, {{1+R}\over 2}\,{e^{-2tH_{0}}}\right)\cr 
=& {v^{(4)}\over 32 (2\pi)^4}\int_0^\infty 
{dt\over t^3}\, \Bigg\{
{f_3(\tq)^8 + f_4(\tq)^8 - f_2(\tq)^8\over f_1(\tq)^8}\cr
&+ 4{f_4(\tq)^4 f_3(\tq)^4 + 
f_4(\tq)^4 f_3(\tq)^4\over f_1(\tq)^4 f_2(\tq)^4}\Bigg\} }}
This can be simplified to:
\eqn\fracthreesimp{
{v^{(4)}\over 32 (2\pi)^4}\int_0^\infty 
{dt\over t^3}\, 
\Bigg\{ 2{f_4(\tq)^8\over f_1(\tq)^8}
+ 8{f_4(\tq)^4 f_3(\tq)^4\over f_1(\tq)^4 f_2(\tq)^4} \Bigg\} }
showing that it is strictly positive. Thus the force is
attractive. This is as expected, since this amplitude describes the
force between $1_f$ and ${\bar 1}_f$.

Finally we evaluate the last independent amplitude:
\eqn\fracfour{\eqalign{
\int_{0}^{\infty} dl\, \langle \dbar3, + \vert e^{-lH_c}\vert 
{\rm D}3, + \rangle
=& \int_{0}^{\infty} {dt \over 2t}\,\tr_{NS-R}\left({1-(-1)^{F} \over 2}\,
{{1-R}\over 2}\,{e^{-2tH_{0}}}\right) 
\cr
=& {v^{(4)}\over 32 (2\pi)^4}\int_0^\infty 
{dt\over t^3}\, \Bigg\{
{f_3(\tq)^8 + f_4(\tq)^8 - f_2(\tq)^8\over f_1(\tq)^8}\cr
&- 4{f_4(\tq)^4 f_3(\tq)^4 + 
f_4(\tq)^4 f_3(\tq)^4\over f_1(\tq)^4 f_2(\tq)^4}\Bigg\} }}
which simplifies to:
\eqn\fracfoursimp{\eqalign{
&{v^{(4)}\over 32 (2\pi)^4}\int_0^\infty 
{dt\over t^3}\, \Bigg\{
2{f_4(\tq)^8\over f_1(\tq)^8} -
8{2f_4(\tq)^4 f_3(\tq)^4\over f_1(\tq)^4 f_2(\tq)^4}\Bigg\} \cr
&={v^{(4)} \over 16(2\pi)^4}\int_{0}^{\infty}{dt \over t^3}
{f_4(\tq)^8\over f_1(\tq)^8}
\left[1 -4{f_1(\tq)^4 f_3(\tq)^4 \over
f_2(\tq)^4 f_4(\tq)^4}\right] }}
Inserting the identity $f_2(\tq)^2 f_3(\tq)^2 f_4(\tq)^2 = 2$, this
amplitude can be rewritten:
\eqn\fracfourfinal{
{v^{(4)} \over 16(2\pi)^4}\int_{0}^{\infty}{dt \over t^3}
{f_4(\tq)^8\over f_1(\tq)^8}
\left[1 - \theta_3(\tq^2)^4 \right] }
and since $\theta_3 > 1$ for real arguments, it follows that the
integrand is strictly negative, implying that the force between the
$1_f$ and $\bar{1'}_f$ is repulsive. This confirms our claim in a
previous section that the force between an adjacent suspended
brane-antibrane pair is repulsive.

To summarise, we have shown that the force between two fractional
$1_f$ branes, or between $1_f$ and $1_f'$, is zero, as expected. The
force between $1_f$ and $\bar{1}_f$ does not vanish, but is instead
attractive, as one expects for a brane-antibrane pair. Finally, the
force between a $1_f$ brane and a ${\bar{1}}'_f$ is repulsive.

\newsec{Concluding Remarks}

{}From the variety of models considered in this paper, a number of
general physical observations and open questions emerge. The
fascinating structure of brane-antibrane pairs and unstable D-branes
in type II string theories, and their various decay modes, are likely
to be a pointer towards more fundamental structures underlying string
theory. The systems we consider, with intersecting branes, lead to an
even more complex picture of decay modes and interactions.

Combining some of the ingredients we have discussed could lead to
dynamically stable non-BPS brane configurations and associated field
theories. For example, one could imagine putting together parallel
branes, which attract, and adjacent branes, which repel, in a variety
of ways.

Various dual configurations exist with the Borromean property, though
we have not described them here. They presumably merit careful
investigation. The model that we have considered here has Bose-Fermi
degeneracy in its spectrum through one-loop order. A curious
occurrence of Bose-Fermi degeneracy in a very different class of
non-BPS systems was described in Ref.\refs{\gabsen,\tatar}, and it
would be interesting to know if there is any common thread that links
these brane configurations and field theories.

\bigskip\medskip
\noindent{\bf Acknowledgements:}
\bigskip
We would like to thank Atish Dabholkar, Keshav Dasgupta, Nick Dorey,
Kimyeong Lee, Soo-Jong Rey, Ashoke Sen, Sandip Trivedi, Gerard Watts
and Piljin Yi for helpful discussions. D.T. would like to express his
gratitude to the Tata Institute of Fundamental Research, the Mehta
Research Institute and the Korean Institute for Advanced Study for
their kind hospitality. D.T. is supported by an EPSRC fellowship.

\listrefs

\end